\documentclass[structabstract]{aa}  
\usepackage{graphicx}
\usepackage[usenames,dvipsnames]{color}
\usepackage{txfonts}
\usepackage{longtable}
\usepackage{supertabular}
\usepackage{natbib}
\defcitealias{Fadda+08}{Paper 0}
\defcitealias{Edwards+10}{Paper 1}
\defcitealias{Edwards-radio}{Paper 2}
\begin{document}

\newcommand{\mic}{$\mu$m}
\newcommand{\lir}{$\rm{L}_{\rm{IR}}$}
\newcommand{\lmir}{$\rm{L}_{24}$}
\newcommand{\mste}{$\rm{M}_{\star}$}

\title{{\em SPITZER} observations of Abell 1763. III. The infrared luminosity 
function in different supercluster environments}
\author{A. Biviano\inst{1}, D. Fadda\inst{2}, F. Durret\inst{3,4},
L.O.V. Edwards\inst{2}, F. Marleau\inst{5}}
\offprints{Andrea Biviano, biviano@oats.inaf.it}

\institute{INAF/Osservatorio Astronomico di Trieste, via G. B. Tiepolo 11, 
I-34131, Trieste, Italy 
\and NASA Herschel Science Center, Caltech 100-22, Pasadena, CA 91125, USA 
\and UPMC Universit\'e Paris 06, UMR 7095, Institut d'Astrophysique de Paris, 
98bis Bd Arago, F-75014 Paris
\and CNRS, UMR 7095, Institut d'Astrophysique de Paris, 98bis Bd Arago,
F-75014 Paris, France
\and   Department of Astronomy and Astrophysics, University of Toronto, 
50 Saint George Street, Toronto, ON M5S 3H4, Canada}

\date{Received / Accepted}

\abstract{The study of galaxy luminosity functions (LFs) in different
  environments provides powerful constraints on the physics of galaxy
  evolution. The infrared (IR) LF is a particularly useful tool since
  it is directly related to the distribution of galaxy star-formation
  rates (SFRs).}{We aim to determine the galaxy IR LF as a function of
  the environment in a supercluster at redshift 0.23 to shed
  light on the processes driving galaxy evolution in and around
  clusters.}{We base our analysis on multi-wavelength data, which
  include optical, near-IR, and mid- to far-IR photometry, as well as
  redshifts from optical spectroscopy. We identify 467 supercluster
  members in a sample of 24-\mic-selected galaxies, on the basis of
  their spectroscopic (153) and photometric (314) redshifts.  IR
  luminosities and stellar masses are determined for supercluster
  members via spectral energy distribution fitting.  Galaxies
    with active galactic nuclei are identified by a variety of methods
    and excluded from the sample. SFRs are obtained for the 432
    remaining galaxies from their IR luminosities via the Kennicutt
  relation.}{We determine the IR LF of the whole supercluster as well
  as the IR LFs of three different regions in the supercluster: the
  cluster core, a large-scale filament, and the cluster outskirts
  (excluding the filament). A comparison of the IR LFs of the three
  regions, normalized by the average number densities of $r$-band
    selected normal galaxies, shows that the filament (respectively,
  the core) contains the highest (respectively, the lowest) fraction
  of IR-emitting galaxies at all levels of IR luminosities, and the
  highest (respectively, the lowest) total SFR normalized by optical
  galaxy richness. Luminous IR galaxies (LIRGs) are almost absent in
  the core region. The relation between galaxy specific SFRs and
  stellar masses does not depend on the environment, and it indicates
  that most supercluster LIRGs are rather massive galaxies with
  relatively low specific SFRs.  A comparison with previous IR LF
  determinations from the literature confirms that the mass-normalized
  total SFR in clusters increases with redshift, but more rapidly than
  previously suggested for redshifts $\la 0.4$.}{The IR LF shows an
  environmental dependence that is not simply related to the local
  galaxy density. The filament, an intermediate-density region in the
  A1763 supercluster, contains the highest fraction of IR-emitting
  galaxies.  We interpret our findings within a possible scenario for
  the evolution of galaxies in and around clusters.}

\keywords{Galaxies: luminosity function - Galaxies: clusters: general - Galaxies: clusters: Abell 1763 - Galaxies: evolution - Galaxies: starburst}

\titlerunning{A1763 IR luminosity function}
\authorrunning{Biviano et al.}

\maketitle

\section{Introduction}
\label{s:intro}
The distribution of galaxy luminosities, i.e. the galaxy luminosity
function (LF), and its environmental dependence have often been used
to provide strong constraints on theories of galaxy evolution
\citep[see e.g. the investigations
  of][]{Zucca+09,Merluzzi+10,Peng+10}. Galaxy environment can be
important in shaping several galaxy properties, such as colors,
morphologies, and star-formation rates \citep[SFRs; see e.g.][for
  reviews]{Biviano08,Gavazzi09}. Since galaxy SFRs are strictly
related to their total infrared (IR) emission \citep{ken98}, powerful
constraints on how galaxies evolve in relation to their environment
are expected to be obtained from the analysis of the galaxy IR LFs.

Following early studies of the IR properties of cluster galaxies with
{\em IRAS} and {\em ISO} \citep[see][and references therein]{MFB05},
the {\em Spitzer Space Telescope} \citep{Werner+04}, the {\em AKARI}
satellite \citep{Murakami+07}, and now the {\em Herschel Space
  Observatory} \citep{Pilbratt+10}, have only recently allowed a
precise derivation of galaxy IR LFs at various redshifts and in
various environments in a precise way.

Most determinations of galaxy IR LFs in cluster and supercluster
environments have so far been based on {\em Spitzer} data.
\citet{Bai+06,Bai+09} analyzed the IR LFs of the rich nearby clusters
Coma and A3266.  According to their analysis, the bright end of the IR
LF has a universal form for local rich clusters, and cluster and field
IR LFs have similar values of their characteristic luminosities,
$\rm{L}_{\rm IR}^{\star}$.  Nearby rich clusters have a lower
star-forming galaxy fraction than field galaxies, although this
fraction increases with cluster-centric distance.

These results were confirmed by \citet{Finn+10}, who analyzed a larger
sample of clusters in the redshift range \mbox{$0.4 \leq \rm{z} \leq
  0.9$}. They noted the similarity in the shape of the cluster and
field IR LFs, and confirmed that there is an increase in the fraction
of luminous IR galaxies (LIRGs\footnote{LIRGs are galaxies with a
  total (8--1000 \mic) IR luminosity \mbox{\lir~$\geq 10^{11} \,
    \rm{L}_{\odot}$}.}) with cluster-centric distance, out to 1.5
virial radii\footnote{The cluster virial radius, $\rm{r}_{200}$, is
  the radius within which the enclosed average mass density of a
  cluster is 200 times the critical density.  The circular velocity
  $\rm{v}_{200}$ is defined as $\rm{v}_{200}=10 \, \rm{H(z)} \,
  \rm{r}_{200}$.  The virial mass $\rm{M}_{200}$ follows from the two
  previous quantities, $\rm{M}_{200}=r_{200} \rm{v}_{200}^2/G$}, where
it is still below the field value. \citet{TDI09} also noted the
absence of LIRGs from the central regions of a sample of 32 X-ray
selected clusters.  Recent {\em Herschel} observations provide
evidence of a lack of IR galaxies not only at the bright end but also
the faint end of the IR LF of the nearby Virgo cluster relative to the
field \citep{Davies+10}.

The environmental dependence of the fraction of high-SFR galaxies may
not be a simple function of cluster-centric distance.  \citet[][Paper
  0]{Fadda+08} detected a large-scale filament\footnote{In
  \citetalias{Fadda+08}, we originally identified two filaments,
  running almost parallel in projection in the sky, but slightly
  separated along line-of-sight velocity space. Subsequent
  spectroscopic observations indicate that the two filaments merge
  into one at large distances from the A1763 cluster core. For
  simplicity, we therefore here refer to a single filament in the
  supercluster.} in the IR, connecting a rich and a poor cluster at
$\rm{z} \sim 0.2$. They observed that the fraction of high-SFR
galaxies is largest in the filament, i.e. larger than in the
cluster core, but also larger than in other, lower density, regions of
the supercluster. The filament detected by \citet{Fadda+08} was the
first to be found via IR observations (with {\em Spitzer}); {\em
  Herschel} observations have recently revealed other large-scale
structure filaments traced by IR-emitting galaxies
\citep{Haines+10,Pereira+10}, but the analysis of their galaxy
populations is still ongoing.

\citet{Koyama+08,Koyama+10} observed that the medium- and low-density
regions of another (more distant, $\rm{z} \sim 0.8$) supercluster host
comparable fractions of star-forming galaxies, while red mid-IR
emitters are preferentially located in medium-density environments,
such as galaxy filaments. Both \citet{Fadda+08} and \citet{Koyama+10}
argued that star-formation is triggered in galaxies in the infall
regions around clusters. \citet{Gallazzi+09} came to the same
conclusion after analyzing the IR galaxy population in a
$\rm{z}=0.165$ supercluster. They also found that while the IR
galaxies prefer to live in medium-density environments, their SFRs are
not particularly high for their stellar masses (\mste), i.e. they have
normal specific SFRs ($\mbox{sSFR} \equiv \rm{SFR}/\rm{M}_{\star}$).

Groups are another environment characterized, as in the case of
filaments, by galaxy densities intermediate between cluster cores and
the field.  \citet{Tran+09} determined the IR LFs of a rich galaxy
cluster and four galaxy groups at $\rm{z} \sim 0.35$. The fraction
of galaxies with a high SFR was found to be four times larger in the
groups than in the cluster, or equivalently, the group IR LF has an
excess at the bright end relative to the cluster IR LF. On the basis
of this result, \citet{Chung+10} interpreted the excess of bright IR
sources in the IR LF of the Bullet cluster ($\rm{z} \sim 0.3$) as
being due to the galaxy population in an infalling group (the
``bullet'' itself).

The IR LF not only depends on the environment, but also on redshift.
\citet{Bai+09} compared the average IR LFs of two nearby ($\rm{z} \leq
0.06$) and two distant ($\rm{z} \sim 0.8$) clusters \citep[using the
  data of][]{Bai+06,Bai+07}. They concluded that there is an evolution
with $\rm{z}$ of both the characteristic luminosity $\rm{L}_{\rm
  IR}^{\star}$ and the normalization of the LF, $\rm{n}^{\star}$, such
that higher-$\rm{z}$ clusters contain more and brighter IR galaxies.
This evolution of the cluster IR LF results in a rapid increase with
$\rm{z}$ in the total SFR of cluster galaxies divided by the total
cluster mass, \mbox{$\Sigma$SFR/mass $\propto (1+\rm{z})^{5.3}$}, a
result anticipated by \citet{Geach+06}, who suggested an even faster
evolution. Another way to characterize this evolution is to look at
the fraction of IR-emitting galaxies (above a given IR luminosity,
\lir) as a function of $\rm{z}$.  This fraction is observed to
increase with $\rm{z}$, a phenomenon called ``the IR Butcher-Oemler
effect'' \citep{STH08,Haines+09b,TDI09}, since it is reminiscent of
the increasing fraction of blue cluster galaxies with $\rm{z}$
\citep{BO84}.  The increasing fraction of LIRGs with $\rm{z}$ appears
however to be a common phenomenon in cluster and field environments
\citep{Finn+10}.

To shed light on the physical processes responsible for the
environmental and redshift dependence of the IR LF, we present a study
of the IR LF of galaxies in the $\rm{z}=0.23$ A1763--A1770
supercluster. Our analysis is restricted to the part of the
supercluster that includes the rich cluster A1763, part of the
filament connecting the two clusters \citepalias[see][]{Fadda+08}, and
the outskirts region around the A1763 cluster core, excluding the
filament itself (see Sect.~\ref{s:irlfenv}).

In Sect.~\ref{s:sample}, we describe our observational data-set
(Sect.~\ref{s:obs}), assign supercluster memberships to the observed
IR galaxies (Sect.~\ref{s:members}), and determine their total IR
luminosities (Sect.~\ref{s:lir}) and stellar masses
(Sect.~\ref{s:stmass}).  In Sect.~\ref{s:irlf}, we describe the
corrections applied to the IR galaxy counts (Sect.~\ref{s:corr}) to
determine the supercluster IR LF (Sect.~\ref{s:sclirlf}). We then
determine the corrected IR LFs of three different regions of the A1763
supercluster to explore environmental effects (Sect.~\ref{s:irlfenv}).
We compare our results with previous results from the literature in
Sect.~\ref{s:comp}. In Sect.~\ref{s:disc}, we discuss our results and
summarize them in Sect.~\ref{s:summ}.

We adopt $\rm{H}_0=70$ km~s$^{-1}$~Mpc$^{-1}$, $\Omega_m=0.3$,
$\Omega_{\Lambda}=0.7$ throughout this paper. In this cosmology,
1 arcmin corresponds to 222 kpc at the cluster redshift.

\section{The data set}
\label{s:sample}
\subsection{Observations}
\label{s:obs}
The data used in this study were obtained as part of a
multi-wavelength observational campaign conducted with several space-
and ground-based telescopes. Details are provided in \citet[][Paper
  1]{Edwards+10}. Here we summarize the main characteristics of the
data set.  A field of $\sim 40 \times 55$ arcmin$^2$ centered on the
A1763 cluster was covered by MIPS 24, 70, and 160~\mic~ observations
from {\em Spitzer}. Two similar fields were also covered by IRAC 3.6,
4.5, 5.8, and 8.0 \mic~ observations from {\em Spitzer.} A similar
area was observed with the Palomar 200~inch telescope in the $r', J,
H,$ and $K_s$ filters. In addition, we obtained spectroscopic
observations for galaxies across the supercluster region, using the
KPNO WIYN and TNG telescopes (paper in preparation). Finally, the
A1763 field was covered by the {\em Sloan Digital Sky Survey} (SDSS
hereafter) in the $u', g', r', i', z'$ photometric bands, and we
collected all data available in the A1763 field from the SDSS Seventh
Data Release (DR7 hereafter). We use Petrosian magnitudes and total
fluxes in the following analyses.

Our sample contains 10876 objects identified at 24~\mic~ in the MIPS
field. The observational technique as well as the depth of our MIPS
observations are very similar to those of the ``verification survey''
in the {\em Spitzer Space Telescope} Extragalactic First Look Survey
\citep[EFLS hereafter;][]{Fadda+06}. For this reason, we assume that
the completeness and purity functions of the EFLS and those of our
survey are identical. This is a conservative assumption because the
EFLS sources were selected using the peak signal-to-noise ratio, while
here sources were selected using the aperture signal-to-noise ratio,
which is more efficient in the rejection of false detections.
Completeness, $\rm{C}_{det}$, is defined as the fraction of real
sources that are detected, and purity, $\rm{P}_{det}$, is defined as
the fraction of real sources among the detected ones.  The
completeness is $\rm{C}_{det} \sim 80$\% at 24~\mic~ flux densities
$\rm{f}_{24}=0.2$ mJy, and close to 100\% at $\rm{f}_{24} > 0.4$ mJy.
The purity is $\rm{P}_{det} \sim 95$\% at $\rm{f}_{24} \geq 0.2$ mJy
and above \citep[see Fig.13 in][]{Fadda+06}.

We base the determination of the IR LFs on the sample of
24~\mic-detected IR-emitting galaxies, since our 70 \mic~ and 160
\mic~ observations are not as deep. About 60\% of the 24~\mic-selected
objects have $\rm{f}_{24} \geq 0.2$ mJy and therefore belong to the
sample with $\geq 80$\% completeness and $\sim 95$\% purity. We use
these completeness and purity estimates in the construction of the
supercluster IR LF (see Sect.~\ref{s:corr}).

\subsection{Supercluster membership}
\label{s:members}

\begin{figure}
\begin{center}
\begin{minipage}{0.5\textwidth}
\resizebox{\hsize}{!}{\includegraphics{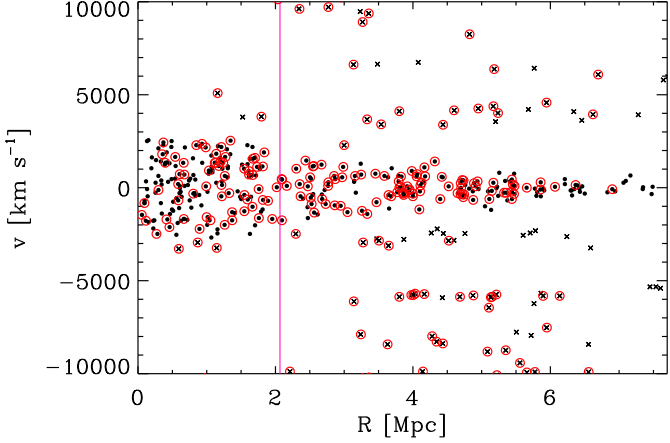}}
\end{minipage}
\end{center}
\caption{The rest-frame velocities versus cluster-centric distances of
  the galaxies with available $\rm{z}$. The vertical solid pink line
  marks the distance to the cluster virial radius,
  $\rm{r}_{200}$. Black dots represent the 357 galaxies selected as
  supercluster members by the algorithm of \citet{Mamon+10},
  interlopers are marked by X's.  Red circles identify 24 \mic~
  emitters. 153 of them are selected as supercluster members.}
\label{f:rv}
\end{figure}

To define the supercluster membership of the galaxies in the cluster
field, we use both spectroscopic ($\rm{z}$) and photometric redshifts
($\rm{z}_p$).

We use the algorithm of \citet{Mamon+10} to identify the supercluster
members among the galaxies with available z. This algorithm tries to
infer the galaxy cluster membership from the location of the galaxy in
the cluster-centric distance -- velocity diagram shown in
Fig.~\ref{f:rv}, based on the modeling of the mass and anisotropy
profiles of cluster-sized halos extracted from a cosmological
numerical simulation. The procedure is more effective than traditional
approaches \citep[e.g.][]{YV77} in rejecting interlopers, while still
preserving cluster members.

The galaxy rest-frame velocities with respect to the cluster mean
velocity are obtained from the usual relation \mbox{$\rm{v} = c
(z-\overline{z}) / (1+ \overline{z})$} \citep{HN79}, where
\mbox{$\overline{\rm{z}}=0.2314$} is obtained using the biweight estimator
\citep{BFG90}. We then obtain the galaxy projected distances from the
cluster center, defined by its X-ray peak emission,
RA=13$^h$35$^m$17.96$^s$, $\delta$=40$^{\circ} 59' 55.8''$
\citep{Cavagnolo+09}.

The algorithm of \citet{Mamon+10} requires initial estimates of the
virial radius, $\rm{r}_{200}$, and circular velocity, $\rm{v}_{200}$,
which we obtain from the cluster velocity dispersion estimate of
\citetalias{Fadda+08}, by following \citet[][Appendix A]{MM07}, and
using the relation of \citet{Gao+08} to infer the concentration of the
cluster mass-density distribution.

We run the procedure on the whole sample of 1364 objects with
available redshift estimates in the supercluster field.  The procedure
is run iteratively until convergence on the number of selected
members. We identify 357 supercluster members (they are shown as
filled dots in Fig.~\ref{f:rv}). Other algorithms
\citep[e.g.][]{dHK96,Fadda+96} lead to very similar membership
definitions.  The average cluster redshift and velocity dispersion
determined for this sample of supercluster members are
$\overline{\rm{z}}=0.2315 \pm 0.0003$ and
$\sigma_{\rm{v}}=1051_{-54}^{+51}$~km~s$^{-1}$. We use these values to
estimate the cluster virial radius and circular velocity as before,
finding $\rm{r}_{200}=2.066$~Mpc and $\rm{v}_{200}=1623$~km~s$^{-1}$,
which do not differ significantly from the initially adopted values.

Of the 357 identified supercluster members, 153 are 24 \mic-emitters.

\begin{figure}
\begin{center}
\begin{minipage}{0.5\textwidth}
\resizebox{\hsize}{!}{\includegraphics{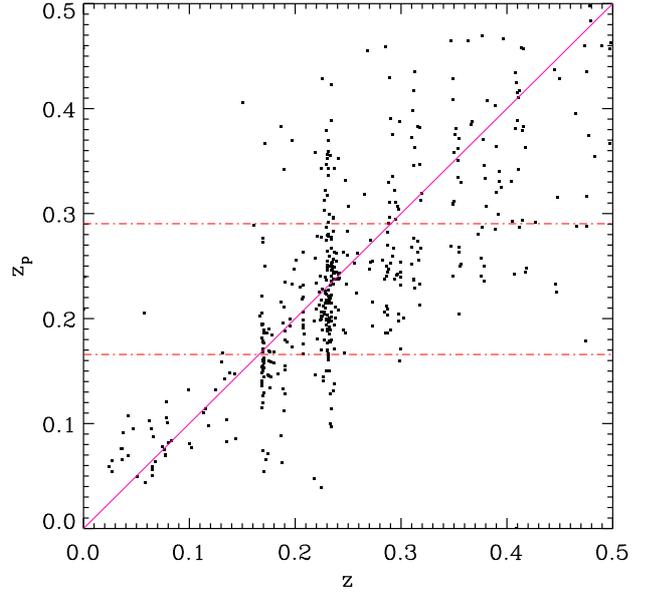}}
\end{minipage}
\end{center}
\caption{SDSS DR7 photometric redshift $\rm{z}_p$ \citep[Artificial
    Neural Network estimates,][]{Oyaizu+08} versus spectroscopic
  redshift $\rm{z}$ for the IR-emitting galaxies with available z and
  $\rm{z}_p$ in the supercluster field (471 galaxies in the displayed
  z and $\rm{z}_p$ ranges). The solid pink line is the identity
  relation $\rm{z}=\rm{z}_p$. The dash-dotted red lines indicate the
  chosen $\rm{z}_p$ range for membership selection (see text and
  Fig.~\ref{f:zp_mem}).}
\label{f:zpz}
\end{figure}

\begin{figure}
\begin{center}
\begin{minipage}{0.5\textwidth}
\resizebox{\hsize}{!}{\includegraphics{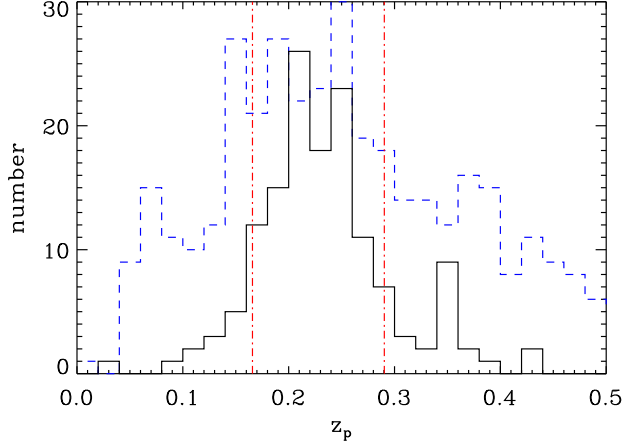}}
\end{minipage}
\end{center}
\caption{The $\rm{z}_p$ distributions for the IR-emitting
  galaxies with available $\rm{z}$ in the supercluster field. The
  solid black (respectively, dashed blue) histogram represents the
  $\rm{z}_p$ distribution for the galaxies selected as members
  (respectively, not selected as members) on the basis of their
  $\rm{z}$.  The two vertical red dash-dotted lines identify the lower
  and upper $\rm{z}_p$ limits used to identify supercluster members in
  the sample of galaxies without $\rm{z}$ (not shown here).}
\label{f:zp_mem}
\end{figure}

To estimate the supercluster membership for the subset of galaxies
without $\rm{z}$, we rely on $\rm{z}_p$-estimates. We consider six
different $\rm{z}_p$-estimates for the galaxies in our sample.  In
particular, we consider the ANNz \citep{CL04} and EAZY \citep{BvDC08}
algorithms, as well as a $\chi^2$ minimization fitting of the spectral
energy distribution (SED, hereafter) of the galaxies in our sample
using SED model templates from \citet{Polletta+07}. We also consider
the three $\rm{z}_p$ estimates directly available from the SDSS
DR7. Of these six $\rm{z}_p$ estimators, we finally adopt one of those
provided in the SDSS DR7, that based on the Artificial Neural Network
technique \citep{Oyaizu+08}. This estimator provides the tightest
correlation between $\rm{z}$ and $\rm{z}_p$ for the subsample of
galaxies in the A1763 field that have both quantities available (see
Fig.~\ref{f:zpz}).

\begin{figure}
\begin{center}
\begin{minipage}{0.5\textwidth}
\resizebox{\hsize}{!}{\includegraphics{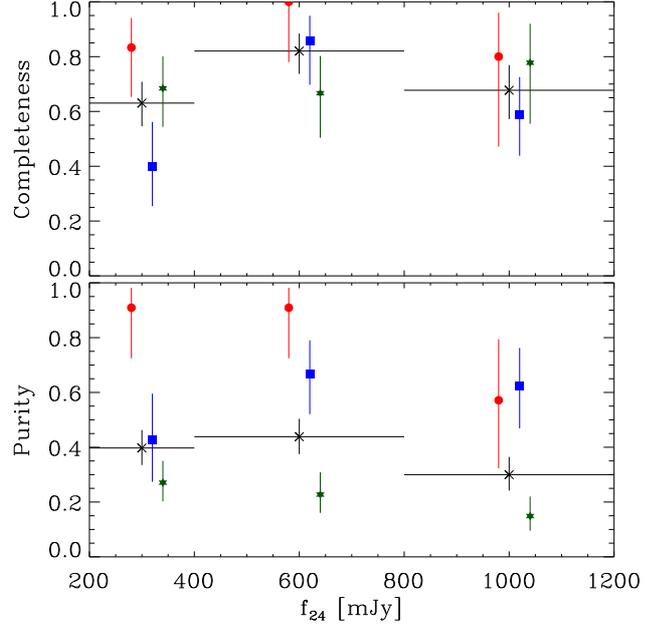}}
\end{minipage}
\end{center}
\caption{Completeness ($\rm{C}_{pm}$) and purity ($\rm{P}_{pm}$) of
  the sample of IR-emitting supercluster members selected on the
  basis of their $\rm{z}_p$ ($0.166 \leq \rm{z}_p \leq 0.290$), as a
  function of $\rm{f}_{24}$. $\rm{C}_{pm}$ and $\rm{P}_{pm}$ are
  estimated using the sample of 24 \mic-emitters with both z and
  $\rm{z}_p$ available, and assuming the members selected on the basis
  of their z are all real members.  The black X's are for the total
  sample. The red dots, blue squares, green stars are for the core,
  filament, and outskirts subsamples, respectively (see
  \ref{s:irlfenv} for the definition of these subsamples).  Horizontal
  bars indicate the $\rm{f}_{24}$ bin intervals.  Vertical bars
  indicate 1-$\sigma$ uncertainties.}
\label{f:cpreg}
\end{figure}

To select the supercluster members in the sample of 24 \mic-emitters
on the basis of their $\rm{z}_p$, we define a $\rm{z}_p$-range around
the mean supercluster redshift. The lower and upper $\rm{z}_p$-limits
that define this selection range must be chosen in such a way as to
maximize the number of real supercluster members with $\rm{z}_p$
within these limits, and, at the same time, minimize the number of
background and foreground galaxies that also happen to have their
$\rm{z}_p$ within these limits. The choice of these $\rm{z}_p$-limits
can only be based on the sample of galaxies with $\rm{z}_p$ {\em and}
z, so that we can perform the most robust $\rm{z}_p$-based membership
selection possible based on the well-established spectroscopic
membership.

We proceed as follows. We assume that the 153 supercluster members
selected on the basis of their $\rm{z}$ are all real members. We then
determine the $\rm{z}_p$-distribution of these 153 galaxies (shown as
a solid black histogram in Fig.~\ref{f:zp_mem}), as well as the
$\rm{z}_p$-distribution of the galaxies with $\rm{z}$ in either the
foreground or the background of the supercluster (dashed blue
histogram in Fig.~\ref{f:zp_mem}). Using the whole sample of galaxies
with $\rm{z}$ and $\rm{z}_p$, we define the purity and completeness to
be, respectively\footnote{For the sake of simplicity hereafter, we use
  the letter ``p'' in lieu of ``$\rm{z}_p$'' in the subscripts.}:
$\rm{P}_{pm} \equiv \rm{N}_{pm \cap zm}/\rm{N}_{pm \cap z}$ and
$\rm{C}_{pm} \equiv \rm{N}_{pm \cap zm}/\rm{N}_{zm \cap p}$, where
$\rm{N}_{zm \cap p}$ is the number of spectroscopically confirmed
cluster members with available $\rm{z}_p$, and $\rm{N}_{pm \cap z}$
(respectively, $\rm{N}_{pm \cap zm}$) is the number of galaxies with z
(respectively, the number of spectroscopically confirmed cluster
members) that have $\rm{z}_p$ within a given $\rm{z}_p$-range.
Following \citet{Knobel+09} we determine the optimal $\rm{z}_p$ range
by minimizing $\sqrt{(1-\rm{P}_{pm})^2+(1-\rm{C_{pm}})^2}$. The
minimum is obtained for $\rm{C}_{pm}=0.73$ and $\rm{P}_{pm}=0.42$,
corresponding to the $\rm{z}_p$-range 0.166--0.290. The dependence of
$\rm{C}_{pm}$ and $\rm{P}_{pm}$ on $\rm{f}_{24}$ is not very strong
(see Fig.~\ref{f:cpreg}). Among the galaxies without $\rm{z}$, 314
have $\rm{z}_p$ within this range.

In Fig.~\ref{f:zpz}, the two red dashed lines indicate the chosen
$\rm{z}_p$-range. It can be seen that most of the supercluster
galaxies fall in that range, but also many of the galaxies that belong
to two other z-peaks, one at $\rm{z} \sim 0.17$, another at $\rm{z}
\sim 0.29$. We consider whether it is possible to increase the purity
of the sample of $\rm{z}_p$-selected cluster members by identifying
and then removing the galaxy structures responsible for these two
z-peaks. The lower-z peak does not correspond to a concentrated
structure in space. The higher-z peak does seem to correspond, at
least partly, to a spatial concentration of galaxies, located at the
edge of the observed Spitzer field. However, removing the (small)
region corresponding to this (presumed) galaxy concentration from our
analysis has hardly any noticeable effect on the results presented in
this paper.

In total, we select 467 IR-emitting galaxies as supercluster members,
153 on the basis of $\rm{z}$, 314 on the basis of $\rm{z}_p$. We base
the derivation of the supercluster IR LF on both the total sample of
members (the $\rm{z} \cup \rm{z}_p$ sample, hereafter), and the sample
of $\rm{z}$-selected members (the z sample, hereafter; see
Sect.~\ref{s:sclirlf}). Using both samples allows us to check the
influence of possible systematic errors because the $\rm{z} \cup
\rm{z}_p$ sample is affected by significant contamination by non-real
members (low purity), while the z sample is affected by larger
incompleteness than the $\rm{z} \cup \rm{z}_p$ sample.

\subsection{Total infrared luminosities} 
\label{s:lir}
To determine the total IR luminosities (\lir) of the 467 supercluster
members, we fit the galaxy SEDs with two sets of model templates, one
from GRASIL \citep{Silva+98}, the other from
\citet{Polletta+07}. These templates span a wide range of galaxy
types, with different formation redshifts, and were used in
\citetalias{Fadda+08} as well as (in part) in \citet{Biviano+04} and
\citet{Coia+05b,Coia+05a}.  In total, we consider 61 SED templates of
galaxies of different ages and types, belonging to the following five
classes:
\begin{itemize}
\item ETGs, early-type galaxies;
\item SFGs, normal star-forming galaxies;
\item SBGs, starburst galaxies;
\item PSBGs, post-starburst galaxies;
\item AGNs, active galactic nuclei.
\end{itemize}

We find the best-fit templates by comparing the template and observed
fluxes via a $\chi^2$ minimization procedure. To compute the template
fluxes in the observed photometric bands, the templates are redshifted
to the galaxy (photometric or spectroscopic) redshifts and convolved
with the filter response curves. The minimization procedure is run
interactively, allowing, when needed, the eye-rejection of deviant
photometric data in the fits of individual galaxy SEDs.  We finally
determine \lir~ by integrating the best-fit model SEDs over the
8--1000 \mic~ rest-frame wavelength range.

\begin{figure}
\begin{center}
\begin{minipage}{0.5\textwidth}
\resizebox{\hsize}{!}{\includegraphics{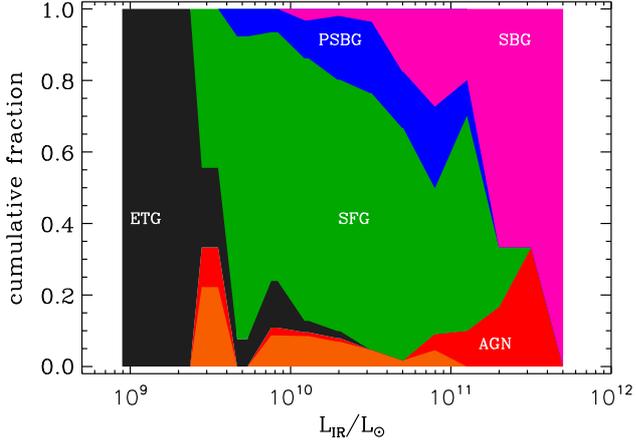}}
\end{minipage}
\end{center}
\caption{The cumulative fractions of galaxies of different SED classes
  as a function of \lir, for the $\rm{z} \cup \rm{z}_p$ sample. The
  pink, blue, green, and black-shaded regions correspond to the
  fractions of SBGs, PSBGs, SFGs, and ETGs, respectively (as
  labeled). The orange and red-shaded regions correspond to the
    fractions of SED-identified AGNs (mostly at low $\rm{L}_{IR}$) and
    AGNs identified in \citetalias{Edwards-radio} from X-ray or radio
    emission, respectively.}
\label{f:fractions}
\end{figure}

\begin{figure}
\begin{center}
\begin{minipage}{0.5\textwidth}
\resizebox{\hsize}{!}{\includegraphics{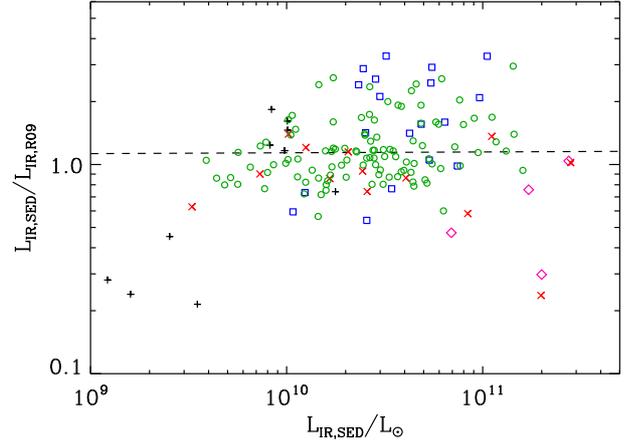}}
\end{minipage}
\end{center}
\caption{Comparison of two IR luminosity estimates, 
  $\rm{L}_{IR,SED}/\rm{L}_{IR,R09}$ versus  $\rm{L}_{IR,SED}$. Different
  symbols identify different galaxy SED classes. Black crosses for
  early-type galaxies (ETGs), red X's for active-galactic nuclei
  (AGNs), pink diamonds for starburst galaxies (SBGs), blue squares
  for post-starburst galaxies (PSBGs), and green circles for normal
  star-forming galaxies (SFGs). The dashed line is the biweight
  average ratio of the sample of non-AGN galaxies.}
\label{f:sed_rieke}
\end{figure}

Given the galaxy IR luminosities, we determine the galaxy SFRs using
the relation of \citet{ken98}, \mbox{$\rm{SFR}
  [\rm{M}_{\odot}/\rm{yr}]= 1.7 \cdot 10^{-10} \cdot
  \rm{L}_{IR}/\rm{L}_{\odot}$}.  This relation is clearly valid only
when a galaxy IR luminosity is not dominated by the emission from an
AGN.  Since most galaxies in our sample lack far-IR photometry, it may
be difficult for us to distinguish AGNs from galaxies with IR emission
dominated by star formation. It is therefore also worth considering
other AGN diagnostics.

In \citet[][Paper 2]{Edwards-radio}, we identified AGNs in the A1763
region using optical, radio, X-ray data, and IRAC colors. Nine of the
AGNs identified in \citetalias{Edwards-radio} are in our sample, and
only one of them has been classified as an AGN based on its SED. This
is unsurprising, since AGNs become visible in different bands at
different stages of their evolution \citep{Hickox+09}, and since the
AGNs identified in \citetalias{Edwards-radio} in the IRAC color
diagram are at the margin of the AGN-identification region
\citepalias[see Fig.~6 in][]{Edwards-radio}. We also adopt the AGN
classification of \citetalias{Edwards-radio} for the 8 galaxies with
non-AGN SED classification, bringing the total of AGNs in our sample
to 35 (13 with available $\rm{z}$). We are therefore confident we have
identified most (if not all) galaxies with AGN-dominated IR emission.

The relative contribution of the different SED classes in
  different \lir~bins is shown in Fig.~\ref{f:fractions} for the
$\rm{z} \cup \rm{z}_p$ sample (the equivalent figure for the z sample
is very similar and not shown here). Fig.~\ref{f:fractions} shows that
SBGs contribute mostly at high \lir, but a significant fraction
of the LIRGs are normal SFGs. The fraction of SFGs and of PSBGs
increases at lower \lir, and SFGs dominate at intermediate
  \lir. Most of the galaxies at the faint-end of the IR LF are
  ETGs. In line with previous results and with our previous analysis
  \citepalias{Edwards-radio}, we find the contribution of AGNs to the
  IR LF of A1763 to be small
  \citep[e.g.][]{Geach+09,Krick+09,Chung+10}, and to increase with
  \lir \citep[e.g.][]{Bothwell+11,Goto+11}.

In order to check the robustness of our SED-based \lir~ estimates we
consider alternative estimates based on direct relations between
$\rm{f}_{24}$ and \lir, from \citet{Rieke+09} and \citet{Lee+10}. When
comparing the different \lir~ estimates, we only consider the
subsample of 140 spectroscopically confirmed non-AGN A1763 members, to
be sure that the comparisons are unaffected by the additional scatter
introduced by photometric redshift errors.  In discussing the results
of these comparisons, we refer to our \lir~ estimates as
$\rm{L}_{IR,SED}$, to the $\rm{f}_{24}$-based \lir~ estimates of
\citet{Rieke+09} as $\rm{L}_{IR,R09}$, and to the $\rm{f}_{24}$-based
\lir~ estimates of \citet{Lee+10} as $\rm{L}_{IR,L10}$.

\begin{figure}
\begin{center}
\begin{minipage}{0.5\textwidth}
\resizebox{\hsize}{!}{\includegraphics{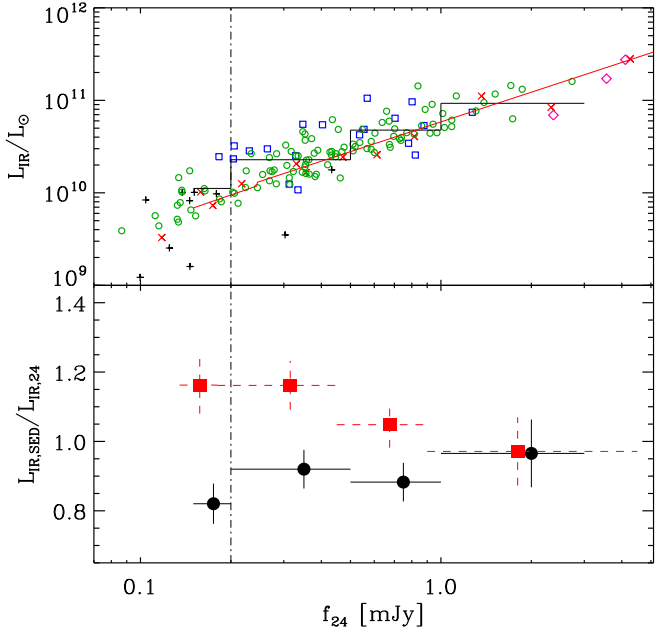}}
\end{minipage}
\end{center}
\caption{{\sl Upper panel:} $\rm{L}_{IR,SED}$ versus $\rm{f}_{24}$ for
  the galaxies of the z sample (spectroscopically confirmed
  supercluster members).  The dash-dotted line indicates the
  completeness limit of 0.2 mJy (see Sect.~\ref{s:sample}). The solid
  red line indicates the relation of \citet{Rieke+09}, and the solid
  black histogram the tabulated values of $\rm{L}_{IR,SED}$ for
  different $\rm{f}_{24}$ from \citet{Lee+10}, scaled from their
  lowest redshift bin to the average redshift of A1763.  The meaning
  of the symbols is the same as in Fig.~\ref{f:sed_rieke}. {\sl Lower
    panel:} ratios of the biweight averages of $\rm{L}_{IR,SED}$ (AGNs
  excluded) to either the tabulated values $\rm{L}_{IR,L10}$ (black
  dots), or the biweight averages of $\rm{L}_{IR,R09}$ (red squares),
  in bins of $\rm{f}_{24}$. In the y-axis label, we generically use
  the notation $\rm{L}_{IR,24}$ to refer to either $\rm{L}_{IR,R09}$
  or $\rm{L}_{IR,L10}$.  Vertical bars are 1$\sigma$ uncertainties in
  the means; horizontal bars indicate the bin ranges. The red squares
  have been slightly displaced along the horizontal axis for clarity.}
\label{f:lir_f24}
\end{figure}

The relation between $\rm{L}_{IR,R09}$ and $\rm{f}_{24}$ is obtained
by combining eqs. (10), (11), (14), and (A6) in \citet{Rieke+09}, and
by interpolating the values of Table~1 in that same paper at the mean
redshift of A1763. Fig.~\ref{f:sed_rieke} shows
$\rm{L}_{IR,SED}/\rm{L}_{IR,R09}$ versus (vs.) $\rm{L}_{IR,SED}$ for our
sample. There is a reasonably good agreement between the two \lir~
estimates, with a rather small systematic offset,
$<\rm{L}_{IR,SED}/\rm{L}_{IR,R09}> = 1.12 \pm 0.05$ \citep[biweight
average, see][]{BFG90}.

\citet{Lee+10} adopted an empirical approach to the \lir~ estimate
from 24~\mic~ flux densities. They stacked 70 and 160 \mic~ images
(taken with {\em Spitzer}) around sources detected at 24 \mic, in
different redshift and $\rm{f}_{24}$ bins.  They then determined \lir~
by fitting the SEDs of the median flux densities in the stacks. To
compare our \lir~ estimates to theirs, we scale their lowest-$\rm{z}$
bin values (at an average redshift $\rm{z}=0.263$, private
communication by N. Lee) to the redshift of A1763, and we estimate the
average \lir~ of our spectroscopically confirmed, non-AGN, A1763
members in the same $\rm{f}_{24}$ bins used by \citet{Lee+10}.

In the upper panel of Fig.~\ref{f:lir_f24}, we show the
$\rm{L}_{IR,SED}$ vs. $\rm{f}_{24}$ value for the galaxies of our z
sample (spectroscopically confirmed supercluster members), as well as
the relations of \citet{Rieke+09} and \citet{Lee+10}.  In the lower
panel of the same figure, we display the ratios of the biweight
averages of $\rm{L}_{IR,SED}$ to either the tabulated values
$\rm{L}_{IR,L10}$ or the biweight averages of $\rm{L}_{IR,R09}$ in
bins of $\rm{f}_{24}$.  It appears that our \lir-estimates are
in-between those obtained using the relations of \citet{Rieke+09} and
\citet{Lee+10}. Overall, these comparisons lend support to the
accuracy of our \lir-estimates.

The small systematic offsets we observe between our $\rm{L}_{IR,SED}$
and either the estimates of \citet{Rieke+09} or those of
\citet{Lee+10} may occur if the SEDs of some galaxies in high-density
regions are not represented by the used model templates, and if they
are atypical of the median SED of the field galaxy population sampled
by \citet{Lee+10}. We note in particular that PSBGs from our z sample
tend to have $\rm{L}_{IR,SED}>\rm{L}_{IR,R09}$, and SBGs
$\rm{L}_{IR,SED}<\rm{L}_{IR,R09}$ (see Fig.~\ref{f:sed_rieke}).
Moreover, we observe that $\rm{L}_{IR,SED}/\rm{L}_{IR,R09}$ increases
with increasing 70 to 24 \mic~ flux density ratio, a correlation that
is significant at the 99\% confidence level. This correlation is
similar to that observed by \citet{Rawle+10} for galaxies in the
Bullet cluster, between the 100 to 24 \mic~ flux density ratio and the
ratio of the SFR obtained from SED fitting, to the SFR obtained from
$\rm{f}_{24}$ via the relations of \citet{Rieke+09}. \citet{Rawle+10}
pointed out that the SFRs obtained from $\rm{f}_{24}$ via the
relations of \citet{Rieke+09} tend to underestimate the true SFRs in
$\sim 40$\% of the cluster galaxies.  We note however that no such
discrepancy exists for field galaxies \citep{Rex+10}, or at least not
for $\rm{z}<0.5$ \citep{Lee+10}.

\begin{figure}
\begin{center}
\begin{minipage}{0.5\textwidth}
\resizebox{\hsize}{!}{\includegraphics{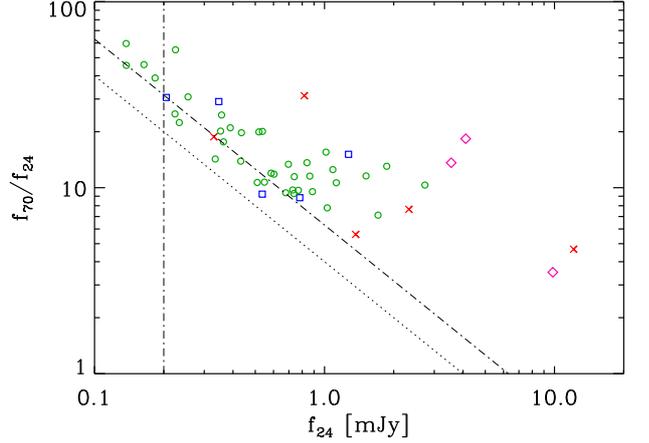}}
\end{minipage}
\end{center}
\caption{The ratio between the 70 and 24~\mic~ flux densities,
  $\rm{f}_{70}/\rm{f}_{24}$ as a function of $\rm{f}_{24}$ for the
  sample of spectroscopically confirmed supercluster members with
  available $\rm{f}_{70}$. The vertical line indicates the
  completeness limit of 0.2 mJy. The two parallel dotted and
    dash-dotted lines indicate the detection and, respectively,
    completeness limit of the 70~\mic~ catalog, 4.0 and 6.3 mJy,
    respectively. See Fig.~\ref{f:sed_rieke} for the meaning of the
  symbols.}
\label{f:f24_f70}
\end{figure}

\citet{Lee+10} found a trend of decreasing $\rm{f}_{160}/\rm{f}_{24}$
with $\rm{f}_{24}$, and a less pronounced trend of
$\rm{f}_{70}/\rm{f}_{24}$ vs. $\rm{f}_{24}$.  They attributed these
trends to an increasing AGN contribution to the IR luminosities of
galaxies with higher 24~\mic~ flux densities. For our sample, there is
an anti-correlation between $\rm{f}_{70}/\rm{f}_{24}$ and
$\rm{f}_{24}$ (see Fig.~\ref{f:f24_f70}), even stronger than the one
observed by \citet{Lee+10}. This anti-correlation may however be
entirely spurious. It may originate from an increasing scatter in the
$\rm{f}_{70}/\rm{f}_{24}$ galaxy colors with decreasing $\rm{f}_{24}$,
combined with the sensitivity limits of our surveys (see dotted and
dot-dashed lines in Fig.~\ref{f:f24_f70}). A similar, albeit smaller,
effect might explain at least part of the anti-correlation seen by
\citet{Lee+10}. As for their interpretation of the anti-correlation,
we note that the galaxies with AGNs do not occupy a special place in
our $\rm{f}_{70}/\rm{f}_{24}$ vs. $\rm{f}_{24}$ diagram (red X's in
Fig.~\ref{f:f24_f70}). Clearer insight into this issue will however
come from our future analysis of the spectral properties of the A1763
supercluster galaxies (paper in preparation).

\subsection{Stellar masses} 
\label{s:stmass}
To determine the galaxy stellar masses, \mste, we fit the SEDs of the
467 supercluster members with the (purely stellar) model templates of
\citet{Maraston05}, adopting the \citet{Kroupa01} initial mass
function and solar metallicity.  We consider only the short-wavelength
parts of the SEDs (rest-frame wavelength $\lambda \leq 4$ \mic), and
allow for dust extinction by modifying the template SEDs according to
the extinction law of \citet{Calzetti+00}, with $\rm{E(B-V)}$ a
parameter free to vary between 0 and 1 \citep{Fontana+04}. On average,
we find that $\rm{E(B-V)}=0.44$ with a dispersion of 0.42.

\begin{figure}
\begin{center}
\begin{minipage}{0.5\textwidth}
\resizebox{\hsize}{!}{\includegraphics{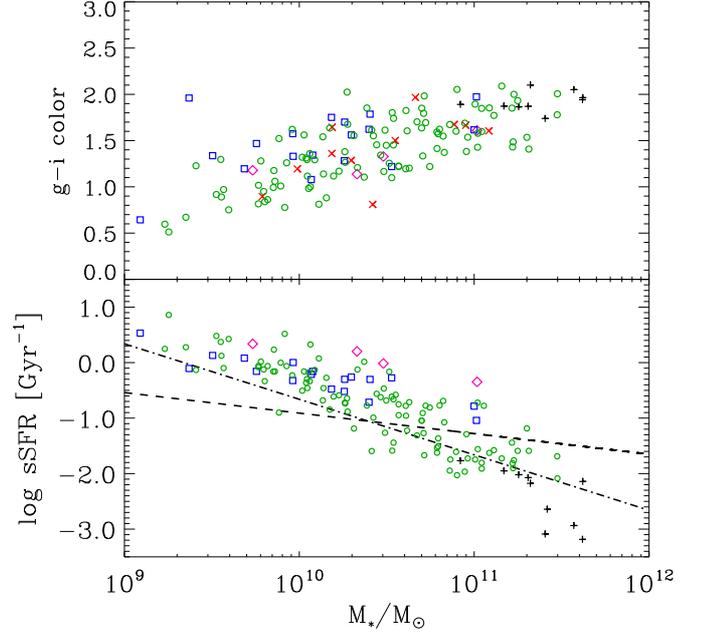}}
\end{minipage}
\end{center}
\caption{{\sl Top:} The $g-i$ color vs. stellar mass for the z sample
  (the spectroscopically confirmed supercluster members). {\sl
    Bottom:} $\log \mbox{sSFR}$ vs. \mste~ for the same z sample,
    AGNs excluded.  The dash-dotted line represent the expected sSFR
  vs. \mste~ relation for a 24 \mic~ source of 0.2 mJy flux density
  (i.e. at the completeness limit of our Spitzer survey), obtained
  using the relations of \citet{Lee+10} and \citet{ken98}. The dashed
  line is the relation of \citet{Oliver+10} for galaxies from the
  Spitzer Wide-area InfraRed Extragalactic Legacy Survey in the
  redshift range $0.2<z<0.3$. In both panels, symbols have the same
meaning as in Fig.~\ref{f:sed_rieke}.}
\label{f:mstar}
\end{figure}

The resulting supercluster galaxy stellar masses are correlated with
the galaxy colors (see Fig.~\ref{f:mstar}, top panel). The correlation
suggests a physical relation between the ages of galaxy stellar
populations and galaxy masses. ETGs have both high \mste~ and red
colors. The scatter in the correlation must be largely intrinsic as it
is not different for galaxies with different values of $\rm{E(B-V)}$.

The galaxy stellar masses \mste~are used to determine the galaxy
specific star formation rates sSFRs ($\rm{sSFR} \equiv
\rm{SFR}/\rm{M}_{\star}$). In the bottom panel of Fig.~\ref{f:mstar},
we display the anti-correlation between \mste~and sSFR in our sample
of spectroscopically confirmed supercluster members, AGNs
excluded. The slope of the correlation is close to $-1$, which is
indicative of an almost flat \mste-\lir~ relation. SBGs have a higher
sSFR per given \mste, relative to other galaxies. This was also found
by \citet{Chung+10} in their study of the Bullet cluster. In addition,
the slope of their sSFR-\mste~ relation is very similar to ours, while
\citet{Oliver+10} found a much flatter relation using a sample of
galaxies from the Spitzer Wide-area InfraRed Extragalactic Legacy
Survey.  This difference is probably related to the way the different
samples were selected, that of \citet{Oliver+10} being closer to a
\mste-selected sample, while our sample is selected on the basis of
the 24~\mic~ flux density. The dashed line in Fig.~\ref{f:mstar}
represents the average expected relation between sSFR and \mste~ for a
24 \mic~ source of 0.2 mJy flux density, corresponding to the limit
below which our sample becomes severely incomplete. This relation has
been obtained using the relation of \citet{Lee+10} between
$\rm{f}_{24}$ and \lir~ at the average redshift of the A1763
supercluster, and the \citet{ken98} relation. Very few sources lie
below the sSFR-\mste~ relation for an $\rm{f}_{24}=0.2$ mJy source,
suggesting that the steeper slope we find for the global sSFR-\mste~
is indeed due to the flux-density limit in our sample.

\section{Infrared luminosity functions}
\label{s:irlf}
\subsection{Completeness and purity corrections}
\label{s:corr}
The determination of the A1763 supercluster IR LF requires the
estimations of the completeness and purity of our sample. We evaluate
three types of completeness and purity corrections, the first for the
source detection, the second for the (spectroscopic or photometric)
redshift determination, the third for the membership assignment.

In the first step, we need to consider the completeness and purity of
the detected 24~\mic~ sources in the photometric catalog, and we model
these corrections following \citet[][see
  Sect.~\ref{s:sample}]{Fadda+06}.  We fit a third order polynomial to
the completeness function of \citet{Fadda+06} to determine the
completeness correction
\begin{equation}
\rm{C}_{det}=1+0.04 \, x + 0.36 \, x^2 + 0.97 \, x^3, 
\end{equation}
for $x \equiv \log \rm{f}_{24} \rm{[mJy]} \leq 0$, and
$\rm{C}_{det}=1$ for $x>0$.  The purity was approximated by a
constant, $\rm{P}_{det} \sim 0.95$ at all flux levels.

In the second step, we consider the completeness of the sample of
sources for which we could establish the cluster membership, i.e. the
sample of sources with either a spectroscopic redshift ($\rm{z}$) or a
photometric redshift ($\rm{z}_p$) estimate.  We call $\rm{N}_{z \cup
  p}$ the number of sources with either $\rm{z}$ or $\rm{z}_p$, and
$\rm{N}$ the total number of sources in the 24~\mic~ catalog.  The
completeness of the sample of sources with either a $\rm{z}$ or
$\rm{z}_p$ estimate is given by $\rm{C}_{z \cup p} \equiv \rm{N}_{z
  \cup p}/\rm{N}$, as a function of $\rm{f}_{24}$.

In the third step, we estimate the completeness and purity of the
sample of selected cluster members (467 in total, see
Sect.~\ref{s:members}) based on their $\rm{z}$ or $\rm{z}_p$.  Since
the membership assignment is imperfect, we need a purity correction to
account for the erroneous membership assignments, and a completeness
correction to account for those real members that have not been
selected.

We first evaluate the membership corrections for the spectroscopic
sample. The fraction of galaxies incorrectly assigned to the cluster
on the basis of their $\rm{z}$ cannot be directly determined from the
data. On the basis of the analyses of cluster-sized halos extracted
from cosmological simulations \citep{Biviano+06,Wojtak+07,Mamon+10},
we assume a membership purity $\rm{P}_{zm}=0.8$ and no completeness
correction for the sample of spectroscopic members.

We then consider the corrections to be applied to the sample of
galaxies without available z, whose membership can only be established
from their $\rm{z}_p$. We proceed in a way similar to that adopted in
Sect.~\ref{s:members} except that now the $\rm{z}_p$-range for
membership selection is fixed to the values previously determined,
$0.166 \leq \rm{z}_p \leq 0.290$. As in Sect.~\ref{s:members}, we have
to determine the completeness and purity by considering galaxies with
z that would qualify as members based on their $\rm{z}_p$, $\rm{N}_{pm
  \cap z}$. A subset of the galaxies in this subsample, $\rm{N}_{pm
  \cap zm}$, are spectroscopic members.  We therefore define the
membership purity of the sample of $\rm{N}_{pm}$ galaxies as the
fraction $\rm{P}_{pm} \equiv \rm{N}_{pm \cap zm}/\rm{N}_{pm \cap z}$
as a function of $\rm{f}_{24}$.  Among the $\rm{N}_{zm \cap p}$
$\rm{z}$-selected members that also have $\rm{z}_p$ estimates, there
are $\rm{N}_{pm \cap zm}$ that would also be identified as members
based on their $\rm{z}_p$.  The membership completeness of the sample
of $\rm{z}_p$-selected members is therefore given by $\rm{C}_{pm}
\equiv \rm{N}_{pm \cap zm}/\rm{N}_{zm \cap p}$ as a function of
$\rm{f}_{24}$.

We define $\rm{N}_{zm}$ to be the number of galaxies defined to be
cluster members based on their $\rm{z}$, and $\rm{N}_{pm}$ the number
of galaxies defined to be supercluster members based on their
$\rm{z}_p$. The corrected number of members is
\begin{equation}
  \rm{N}_c \equiv \frac{\rm{P}_{det}}{\rm{C}_{det}} \cdot
  \frac{1}{\rm{C}_{z \cup p}} \cdot (\rm{P}_{zm} \cdot \rm{N}_{zm} +
  \frac{\rm{P}_{pm}}{\rm{C}_{pm}} \cdot \rm{N}_{pm}).
\label{e:compurp}
\end{equation}

By combining the data for galaxies with available $\rm{z}_p$ with
those for galaxies with available $\rm{z}$, we obtain a larger sample
of members, but at the expense of a larger uncertainty in the
membership assignments. The resulting sample (the $\rm{z} \cup
\rm{z}_p$ sample) is therefore more complete, but less pure than the
sample of supercluster members constructed using only galaxies with
available $\rm{z}$ (the z sample). To check for possible systematics
related to our purity corrections, we also determine the IR LF for the
z sample. We call $\rm{N}_z$ the number of galaxies with available
$\rm{z}$ among the total of 24 \mic-selected sources.  The
completeness of this spectroscopic sample is $\rm{C}_z \equiv
\rm{N}_z/\rm{N}$.  Therefore, the corrected number of members of the z
sample is
\begin{equation}
\rm{N}_{cz} \equiv \frac{\rm{P}_{det}}{\rm{C}_{det}} \cdot \frac{1}{\rm{C}_z} \cdot \rm{P}_{zm} \cdot \rm{N}_{zm}.
\label{e:compurz}
\end{equation}
We note that in eqs.~\ref{e:compurp}, \ref{e:compurz} we have omitted
the explicit $\rm{f}_{24}$-dependence of the individual terms to
simplify the notation.

\subsection{The supercluster luminosity function}
\label{s:sclirlf}
We determine the IR LF of the supercluster by counting the galaxies in
(logarithmic) luminosity bins, and weighting the counts by the
correction functions described above \citep[Sect.~\ref{s:corr}; the
  same procedure was used by][]{Rujopakarn+10}. Since the correction
factor becomes very high at low fluxes, we only consider galaxies with
$\rm{f}_{24} \geq 0.2$ mJy (317 out of the originally selected 467
cluster members, 124 selected as members on the basis of their
$\rm{z}$). We multiply the counts by the fractions of non-AGN galaxies
in each \lir-bin (see Fig.~\ref{f:fractions}) to remove the AGN
contribution from the IR LF.

We obtain two determinations of the IR LF by using in one case the
$\rm{z} \cup \rm{z}_p$ sample, and in the other case the z sample (see
Sect.~\ref{s:members}). The error bars of the IR LF are estimated with
a bootstrap re-sampling technique \citep{ET86}.  Both the galaxy
counts and the correction functions are computed for each bootstrap
re-sampling.

\begin{figure}
\begin{center}
\begin{minipage}{0.5\textwidth}
\resizebox{\hsize}{!}{\includegraphics{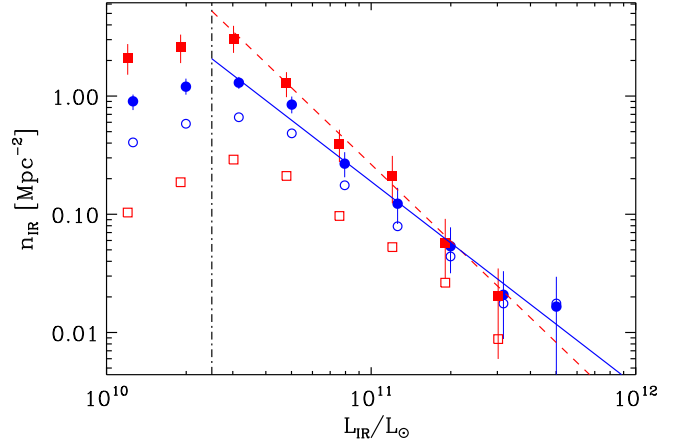}}
\end{minipage}
\end{center}
\caption{The IR LF of A1763.  Filled and empty symbols represent the
  counts after and, respectively, before purity and completeness
  corrections. Counts have been multiplied by the fractions of
    non-AGN galaxies in each \lir-bin (see Fig.~\ref{f:fractions}) to
    remove the AGN contribution from the IR LF.  Blue dots
  (respectively, red squares) represent the counts based on the
  $\rm{z} \cup \rm{z}_p$ (respectively, z) sample.  1$\sigma$ error
  bars based on 100 bootstrap re-samplings are shown.  The square
  symbols have been displaced by $-0.02$ in $\log \rm{L}_{IR}$ for
  clarity.  The vertical dash-dotted line indicates the \lir~ lower
  limit ($\rm{L}_{IR}/\rm{L}_{\odot}=2.5 \cdot 10^{10}$) that
  corresponds to the 0.2 mJy flux density limit adopted for the
  determination of the IR LF. The solid blue (respectively dashed red)
  line represents the power-law best-fit to the IR LF represented by the
  blue filled dots (respectively, red squares) with
  $\rm{L}_{IR}/\rm{L}_{\odot} \geq 2.5 \cdot 10^{10}$.}
\label{f:irlf}
\end{figure}

The resulting IR LF determinations are shown in Fig.~\ref{f:irlf}.
Filled symbols represent the corrected counts, open symbols the
uncorrected counts, and the ratios of the two give the correction
factors applied (based on eqs.~\ref{e:compurp} and \ref{e:compurz} for
the $\rm{z} \cup \rm{z}_p$ and z sample, respectively). The two
determinations agree within the error bars down to
$\rm{L}_{IR}/\rm{L}_{\odot} \simeq 4 \cdot 10^{10}$; at lower \lir~
the correction factor for the counts in the z sample is very large
($>10$), and therefore rather uncertain. The agreement of the two IR
LF determinations down to $\rm{L}_{IR}/\rm{L}_{\odot} \simeq 4 \cdot
10^{10}$ suggests that the completeness and purity corrections that we
have applied to the two subsamples are reasonably accurate.

\begin{table}
\centering
\caption{Slope parameters from power-law fits to the IR LFs}
\label{t:fits}
\begin{tabular}{ccccc}
\hline
Sample & Region & $\alpha$ & $\chi^2$ & dof \\
\hline
& & & & \\
$\rm{z} \cup \rm{z}_p$ & Whole     & $-1.7 \pm 0.1$ & 3.3 & 6 \\ 
$\rm{z} \cup \rm{z}_p$ & Core      & $-2.2 \pm 0.7$ & 0.5 & 4 \\ 
$\rm{z} \cup \rm{z}_p$ & Filament  & $-1.5 \pm 0.3$ & 3.1 & 5 \\ 
$\rm{z} \cup \rm{z}_p$ & Outskirts & $-2.4 \pm 0.4$ & 8.8 & 6 \\ 
& & & & \\
\hline
& & & & \\
$\rm{z}$               & Whole     & $-2.1 \pm 0.5$ & 0.6 & 5 \\ 
$\rm{z}$               & Core      & $-1.8 \pm 0.6$ & 0.1 & 3 \\ 
$\rm{z}$               & Filament  & $-1.8 \pm 0.3$ & 1.6 & 5 \\ 
$\rm{z}$               & Outskirts & $-2.2 \pm 0.9$ & 0.2 & 3 \\ 
& & & & \\
\hline
\end{tabular}
\end{table}

The vertical dash-dotted line in Fig.~\ref{f:irlf} indicates the \lir~
lower limit corresponding to the adopted limit of $\rm{f}_{24}=0.2$ mJy 
for the IR LF determination, \lir~$\simeq 2.5 \cdot 10^{10} \,
\rm{L}_{\odot}$ (see Fig.~\ref{f:lir_f24}; this limit is not very
precise because of the dispersion in the \lir-$\rm{f}_{24}$ relation).

We try fitting the IR LF at \lir~$\geq 2.5 \cdot 10^{10} \,
\rm{L}_{\odot}$ with a \citet{Schechter76} function, but the best-fit
parameters are poorly constrained. This is mostly because the
Schechter function decreases steeply at high luminosities, beyond
$\rm{L}_{\rm IR}^{\star}$, while our IR LF does not show a change in
slope over the full range of luminosities. Some authors have advocated
the use of a double power-law as a fitting function for IR LFs
\citep{Babbedge+06,Goto+11}. The characteristic luminosity at which
the IR LF of field galaxies changes slope in this case is $\sim 5
\times 10^{10} \, \rm{L}_{\odot}$ \citep{Goto+11}, which is close to
our adopted completeness limit. A single power-law function can thus
be expected to provide a good fit to our IR LF over the full range of
luminosities down to the completeness limit. This is indeed the case,
as shown in Fig.~\ref{f:irlf}, where the solid and dash-dotted lines
represent the best-fit power-law functions for the two samples. The
best-fit values of the slope parameter are given in Table~\ref{t:fits}
(region 'Whole'); the quality of the fits, as indicated by the listed
$\chi^2$ values, is good, and indicates that a two-parameter fit
(e.g. with a Schechter function) is not required (note that the
$\chi^2$ values are {\sl not} reduced $\chi^2$). The values we obtain
for the two samples are compatible within the 1-$\sigma$ error bars.

\subsection{Environmental dependence}
\label{s:irlfenv}
To investigate possible environmental effects on the IR LF, we
consider three different regions of the A1763 supercluster.  To more
clearly define the location of the large-scale filament identified in
\citetalias{Fadda+08}, we determine the galaxy density map of the
supercluster, as traced by IR-emitting, star-forming galaxies, by
running an adaptive-kernel technique \citep[see, e.g.,][]{Biviano+96}
on the sample of 432 non-AGN supercluster members (see
Sect.~\ref{s:members} and \ref{s:lir}). We consider only the $\rm{z}
\cup \rm{z}_p$ sample in this case, because it is more complete than
the z sample, and completeness is more important than purity when
determining the density map, as long as there are no contaminating
background or foreground structures in the sample (and we think there
are not, see Sect.~\ref{s:members}).

\begin{figure}
\begin{center}
\begin{minipage}{0.5\textwidth}
\resizebox{\hsize}{!}{\includegraphics{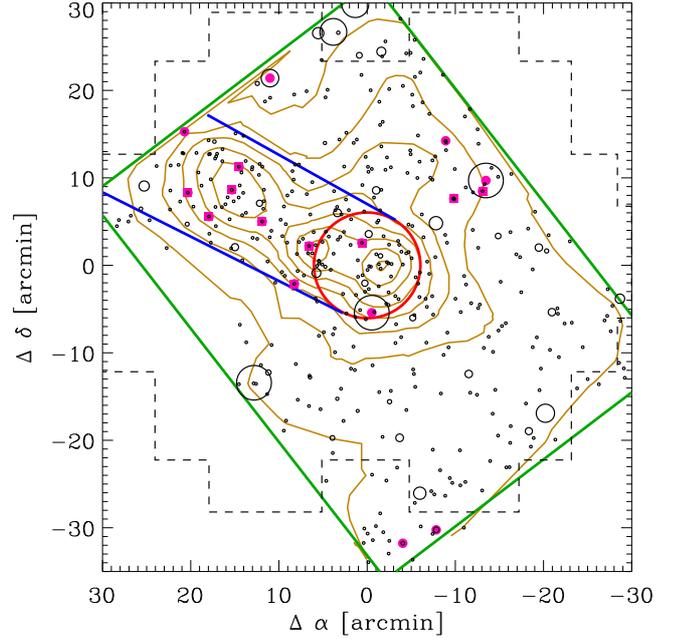}}
\end{minipage}
\end{center}
\caption{Diagram illustrating the three mutually exclusive regions for
  which independent IR LFs have been defined.  The red circle defines
  the ``core'' region, which is centered on the cluster A1763 and
  extends to a radius $\rm{r}_{500}=1.34$ Mpc. The two almost parallel
  blue segments delimit the extended over-density of supercluster
  members outside the core, which we identify with the large-scale
  ``filament'' region, discovered in \citetalias{Fadda+08}. The
  ``outskirts'' region is delimited by the green parallelogram, which
  represents the extent of the 24~\mic~ observations, excluding the
  core and filament regions. We also show the region corresponding to
  the Palomar $r$-band observations (connected dashed black segments).
  The (brown) contours are isocontours of galaxy number density,
  linearly spaced, obtained by running an adaptive-kernel technique on
  the spatial distribution of the 432 non-AGN, IR-emitting
  supercluster members of the $\rm{z} \cup \rm{z}_p$ sample.
  Positions of these galaxies are indicated by the (black) circles,
  with sizes proportional to the galaxy $\mbox{sSFRs}$. Filled (pink)
  symbols mark the positions of the LIRGs in the $\rm{z} \cup
  \rm{z}_p$ sample, squares for spectroscopically confirmed members,
  dots for members selected on the basis of their $\rm{z}_p$.}
\label{f:regions}
\end{figure}

The result is shown in Fig.~\ref{f:regions}. A clear over-density of
galaxies is seen extending to the north-east direction from the
central cluster region\footnote{Follow-up spectroscopic observations
  show that this galaxy over-density continues beyond the region
  covered by our Spitzer observations. The apparent cut-off of the
  filamentary structure visible at the edge of the Spitzer field in
  Fig.~\ref{f:regions} is an edge-effect of the adaptive-kernel
  algorithm.}.  This region coincides with the filamentary
structure(s) found in \citetalias{Fadda+08}. We draw two almost
parallel lines delimiting this over-density region in order to
identify the ``filament'' region. We clearly define the ``filament''
region by excluding the ``core'' region, which we define to be the
1.34 Mpc circular region centered on the A1763 cluster center.  This
region corresponds to the projection of the sphere with a mass
over-density 500 times the critical density, and its radius is
estimated as $r_{500}=0.65 \, r_{200}$, using the mass profile of
\citet{NFW97} with a concentration parameter $c=4$, typical of massive
galaxy clusters \citep[e.g.][]{KBM04}.  We finally define the
``outskirts'' region as the remaining part of the observed 24~\mic~
survey region, excluding the core and the filament.

\begin{figure*}
\begin{center}
\begin{minipage}{1.\textwidth}
\resizebox{\hsize}{!}{\hbox{\includegraphics{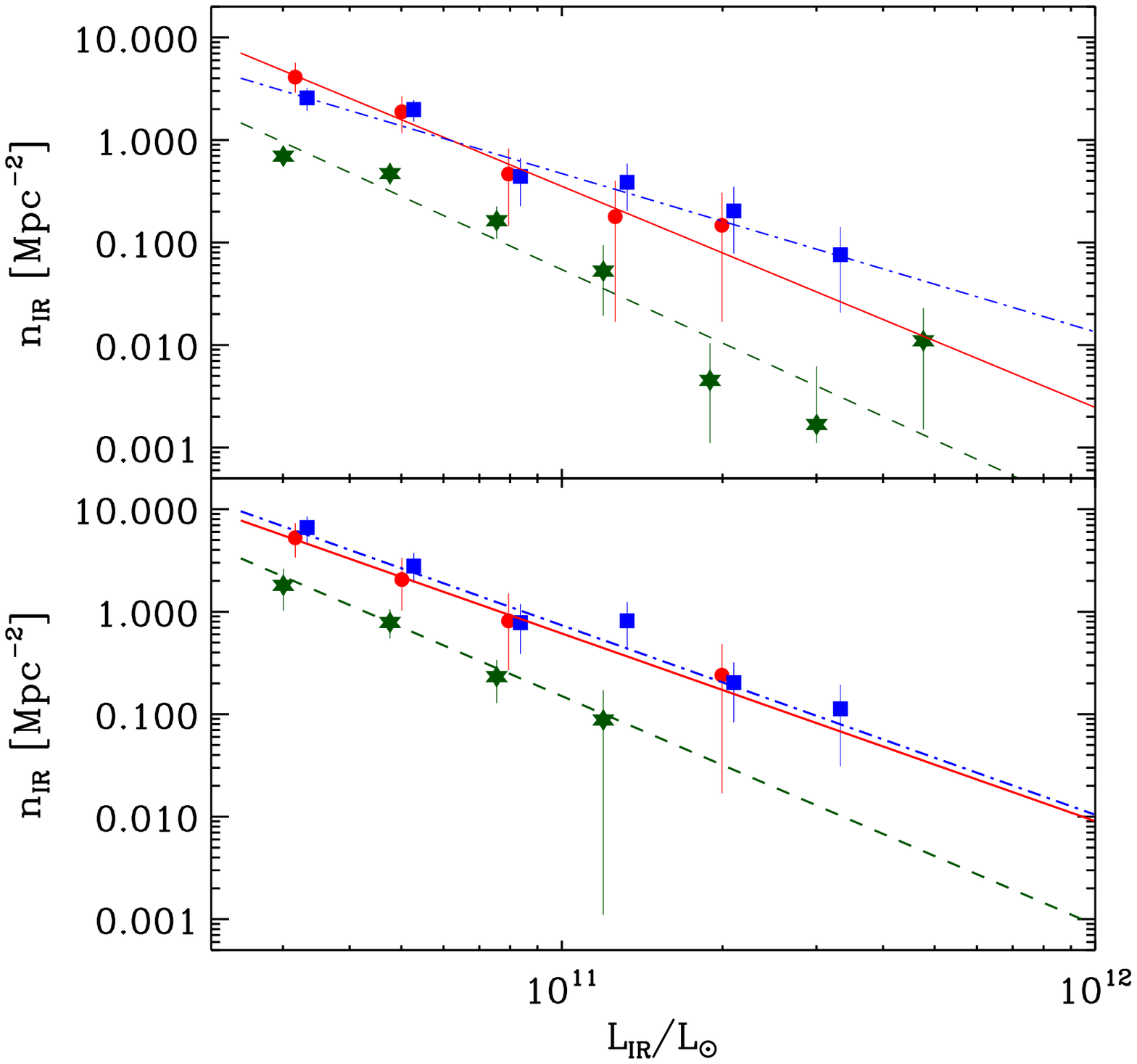} \includegraphics{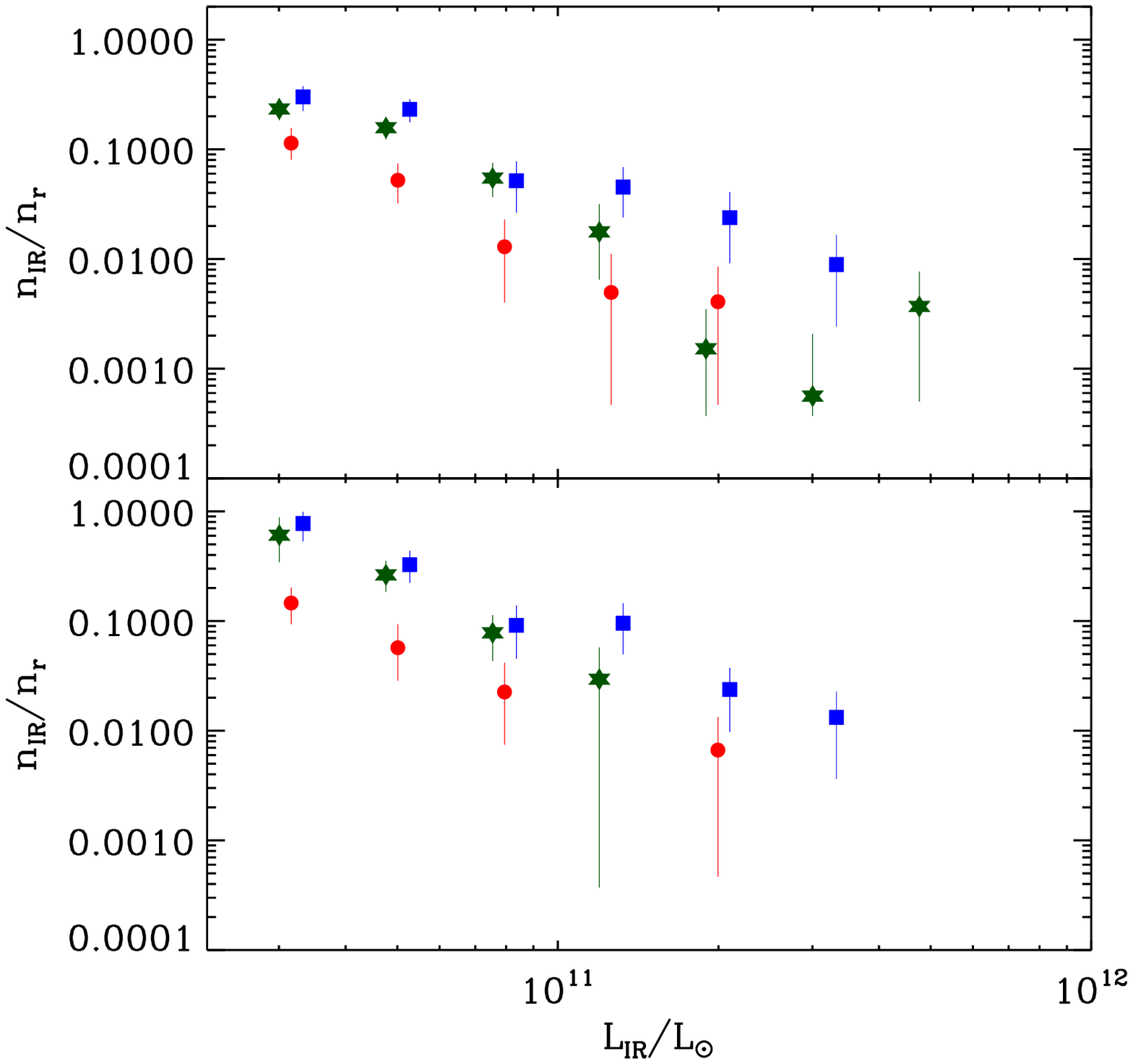}}}
\end{minipage}
\end{center}
\caption{The IR LFs of three different regions in the A1763
  supercluster: core (red dots), filament (blue squares), and
  outskirts (green stars). 1$\sigma$ error bars based on 100 bootstrap
  re-samplings are shown. All IR LFs are corrected for purity and
  completeness in the corresponding regions. Counts have been
  multiplied by the fractions of non-AGN galaxies in each \lir-bin and
  in each region to remove the AGN contribution from the IR LFs. The
  IR LFs of the upper (respectively, lower) panels have been obtained
  using the $\rm{z} \cup \rm{z}_p$ sample (respectively, z sample) of
  supercluster members.  {\sl Left panels} show the number densities
  of IR-emitting galaxies. Lines represent the best-fit power-law
  functions to the IR LFs. In the {\sl right panels}, the number
  densities of IR galaxies have been normalized by the average number
  densities of galaxies with a $r$-band luminosity $\geq 7 \, 10^9 \,
  \rm{L}_{\odot}$ within each region.}
\label{f:3irlf}
\end{figure*}

The IR LFs of the three different regions were determined as described
in Sect.~\ref{s:corr}, using completeness and purity corrections that
are appropriate for each considered region, and multiplying the counts
by the fractions of non-AGN galaxies in each \lir-bin and each region
to remove the AGN contribution from the IR LFs.  Error bars were
determined via a bootstrap re-sampling procedure. The three IR LFs are
displayed in the left-hand panels of Fig.~\ref{f:3irlf}, for both the
$\rm{z} \cup \rm{z}_p$ (top panel) and the z sample (bottom
panel). Power-law function fits to the three IR LFs are shown as
dashed lines, and the best-fitting values of the slope parameter are
given in Table~\ref{t:fits}.

The slopes of the three region IR LFs do not differ significantly, but
taken at face value they suggest that the filament has a flatter IR LF
than both the outskirts and (for the $\rm{z} \cup \rm{z}_p$ sample)
the core. The IR LF of the filament region is flatter because of an
excess of LIRGs relative to the other regions. This is also apparent
from a visual inspection of Fig.~\ref{f:3irlf} and also of
Fig.~\ref{f:regions}, where we show the spatial positions of all
supercluster members in the $\rm{z} \cup \rm{z}_p$ sample and indicate
the LIRGs with pink symbols (square symbols for LIRGs of the z
sample).

Fig.~\ref{f:3irlf} (left panels) also shows that at lower \lir, the
number densities of IR-emitting galaxies are similar in the core and
in the filament regions, and lowest in the outskirts region. When
considering the implications of this comparison, one must take into
account that the three selected regions are characterized by different
densities of normal galaxies, highest in the core, lowest in the
outskirts. Similarities in the IR LFs of different regions could be
caused by a combination of different densities of normal galaxies and
different fractions of IR-emitting galaxies among the total.
Viceversa, different IR LFs could simply reflect differences in the
densities of normal galaxies combined with similar IR-emitting galaxy
fractions among the total.

It is therefore also important to compare the relative fractions of
IR-emitting galaxies in the different regions. For this, we must
determine the densities of normal galaxies in the three different
regions. By adopting the same methodology used for a derivation of the
IR LF (see Sect.~\ref{s:corr}), we determine the $r$-band LFs in the
three regions. These LFs are well fitted by Schechter functions, and
their shapes are not statistically different according to a $\chi^2$
test \citep{DeGroot87}.  We then integrate these LFs to derive the
number densities of $r$-band selected galaxies with $r$-band
luminosity\footnote{Because of the similar shapes of the $r$-band LFs
  of the three regions, the exact choice of the luminosity limit for
  the integration of the $r$-band LFs does not strongly affect the
  relative ratios of the three regions number densities ($\leq \pm
  10$\% when the luminosity limit is increased by up to a factor
  three).} $\rm{L}_r \geq 7 \, 10^9 \, \rm{L}_{\odot}$.  This
luminosity represents the lower limit above which our determinations
of the $r$-band LFs appear to be robust, i.e. independent of sample
choice (the $\rm{z} \cup \rm{z}_p$ sample or the z sample). It
corresponds to a stellar mass \mste $\approx 6$--$7 \, 10^9 \,
\rm{M}_{\odot}$ \citep{Bell+03,Bernardi+10}, which roughly matches the
lower stellar mass limit of the IR detected galaxy population in A1763
(see Fig.~\ref{f:mstar}).

The $r$-band number densities ($\rm{n}_r$) are given in
Table~\ref{t:sfr}. The number density of $r$-band selected galaxies in
the filament is in-between those of the core and the outskirts, as
expected.

\begin{figure}
\begin{center}
\begin{minipage}{0.5\textwidth}
\resizebox{\hsize}{!}{\includegraphics{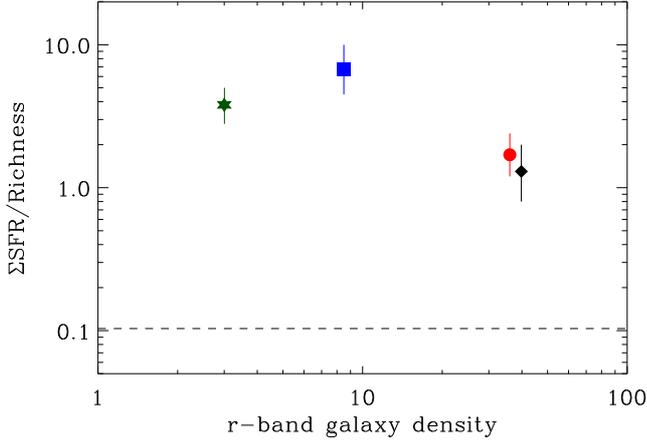}}
\end{minipage}
\end{center}
\caption{$\Sigma$SFR/richness (in units of $\rm{M}_{\odot} \,
  \rm{yr}^{-1}$) as a function of the average $r$-band galaxy density
  within each region.  Richnesses and densities are estimated using
  galaxies brighter than $7 \, 10^9 \, \rm{L}_{\odot}$ in the
  $r$-band.  Different symbols identify the three different regions,
  as in Fig.~\ref{f:3irlf}, and the filled black diamond
  identifies the region within $\rm{R} \leq 0.5 \, \rm{r}_{200}$.
  1$\sigma$ error bars are shown.  The dashed line represents the
  expected value at the cluster mean $z$, using the relation of
  \citet{Bai+09} between $\Sigma$SFR/mass and $z$, and the
  richness/mass value of the A1763 virial ($\rm{R}\leq \rm{r}_{200}$)
  region (see Sect.~\ref{s:comp}).}
\label{f:tsfrr}
\end{figure}

We divide the IR LFs of the three regions by their $\rm{n}_r$ to
produce the plots shown in the right-hand panels of Fig.~\ref{f:3irlf}
(top panel: $\rm{z} \cup \rm{z}_p$ sample; bottom panel: z sample).
When scaled by the relative densities of $r$-band selected cluster
members in the different regions, the filament displays the highest
over-density of IR-emitting galaxies, with respect to both the core
and the outskirts, at all \lir. According to a $\chi^2$ test
\citep{DeGroot87}, the difference is very significant with respect to
the core ($99.9$~\% significance level for both the $\rm{z} \cup
\rm{z}_p$ and the z sample), but not significant with respect to the
outskirts. The global difference between the outskirts and the core IR
LFs is marginally significant (98~\% significance level for both the
$\rm{z} \cup \rm{z}_p$ and the z sample).

\begin{table}
\centering
\caption{Properties of different supercluster regions. $\Sigma$SFR is
  the total SFR of IR-emitting galaxies. $\Sigma$SFR/M$_{200}$ is the
  $\Sigma$SFR normalized by the total gravitational cluster mass.}
\label{t:sfr}
\begin{tabular}{lcccc}
\hline
& & & & \\
Property & \multicolumn{4}{c}{Supercluster regions} \\
& core & filament & outskirts & $\rm{R} \leq 0.5 \, \rm{r}_{200}$ \\
\hline
& & & & \\
Area & 5.9 & 17.4 & 90.4 & 3.6 \\

 [Mpc$^2$] & & & & \\
\\
$\rm{n}_r$ & $36.0 \pm 6.7$ & $8.5 \pm 1.8$  & $3.0 \pm 0.4$ & $39.8 \pm 7.3$ \\

 [Mpc$^{-2}$] & & & & \\
\\
$\Sigma$SFR/(Area $\cdot \rm{n}_r$) & $1.7_{-0.5}^{+0.7}$ & $6.7_{-2.2}^{+3.3}$ & $3.8_{-1.0}^{+1.2}$ & $1.3_{-0.5}^{+0.7}$ \\

 [\mste yr$^{-1}$] & & & & \\
\\
$\Sigma$SFR/M$_{200}$ & & & & $26_{-9}^{+11}$ \\ 

 [M$_{\odot}$ yr$^{-1}$/$10^{14}$ M$_{\odot}$] & & & & \\

& & & & \\
\hline
\end{tabular}
\end{table}

The difference between the IR LFs in the three supercluster regions
reflects a different SFR per galaxy. By integrating the IR LF down to
our adopted completeness limit, we obtain the total \lir~ of galaxies
in the three regions, which we then convert to a total SFR
($\Sigma$SFR hereafter) via the relation of \citet{ken98}. We divide
the $\Sigma$SFR values of the three regions by the areas of the three
regions and the number densities of $r$-band selected galaxies in the
three regions to obtain the average SFRs per $r$-band selected galaxy
in each region\footnote{These averages are clearly not representative
  of the typical galaxy SFR, as they are biased high by the high SFRs
  in the (relatively few) very bright IR emitting galaxies.}. The
values are given in Table~\ref{t:sfr} for the $\rm{z} \cup \rm{z}_p$
sample (consistent values are found for the z sample, within the
errors); they are displayed as a function of the average density of
$r$-band selected galaxies in Fig.~\ref{f:tsfrr}. The average SFR is
highest for the filament region, intermediate for the outskirts
region, and lowest for the core region.  The difference between the
filament and the core values is significant at slightly more than
2-$\sigma$, i.e. at the 98~\% confidence level for a Gaussian
distribution of errors, that between the core and the outskirts values
is significant at the 96~\% confidence level, and that between the
outskirts and the filament values is not significant ($<90$~\%
confidence level).

\begin{figure}
\begin{center}
\begin{minipage}{0.5\textwidth}
\resizebox{\hsize}{!}{\includegraphics{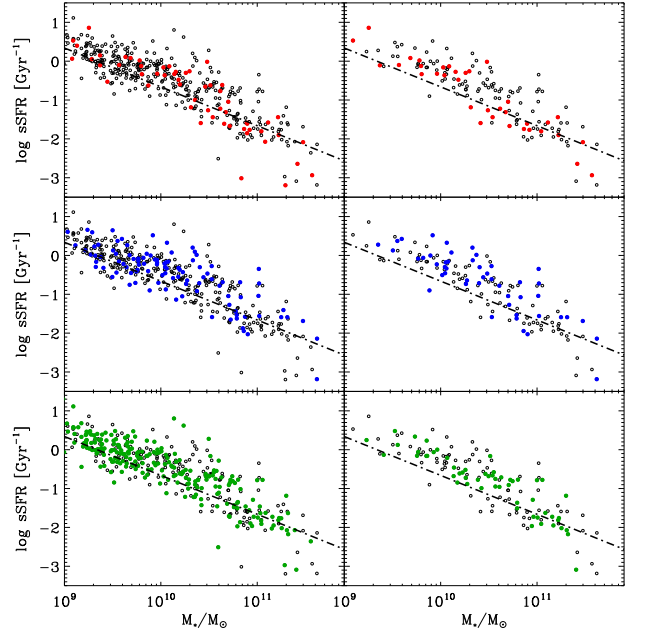}}
\end{minipage}
\end{center}
\caption{Galaxy sSFR vs. \mste~ in three different regions of the
  A1763 supercluster. Open (black) circles represent all non-AGN,
    IR-emitting supercluster members.  Filled dots identify
  supercluster members in the core region (red dots, top panels), in
  the filament region (blue dots, middle panels), in the outskirts
  region (green dots, bottom panels).  The panels on the left are
  based on the $\rm{z} \cup \rm{z}_p$ sample, those on the right on
  the $\rm{z}$ sample. The dash-dotted line has the same meaning
    as in Fig.~\ref{f:mstar}.}
\label{f:ssfrmste}
\end{figure}

Are the excess LIRGs in the filament region massive galaxies or
low-mass galaxies with high levels of sSFRs? To understand this issue,
in Fig.~\ref{f:regions} we use symbol sizes proportional to the galaxy
sSFRs to represent the spatial positions of the 432 non-AGN
supercluster members. Most of the LIRGs have rather low sSFRs, meaning
that they have both high \lir~and high \mste.

Another way to look at this issue is to compare the bi-dimensional
distributions of galaxies in different regions in a sSFR vs. \mste~
diagram, shown in Fig.~\ref{f:ssfrmste} for both the $\rm{z} \cup
\rm{z}_p$ and z sample (left- and right-hand panels, respectively).
Galaxies of different regions of the superclusters appear to have
similar sSFR--\mste~ distributions. A statistical assessment of this
result is obtained by comparing the sSFR--\mste~ distributions two by
two via bi-dimensional Kolmogorov-Smirnov tests
\citep{Peacock83,FF87}, under the null hypothesis that the
distributions are drawn from the same parent one.  Only in one case,
core vs. outskirts, and only for the $\rm{z} \cup \rm{z}_p$ sample we
do find that the null hypothesis is rejected, but only with marginal
significance (97~\% confidence level).

\begin{figure}
\begin{center}
\begin{minipage}{0.5\textwidth}
\resizebox{\hsize}{!}{\includegraphics{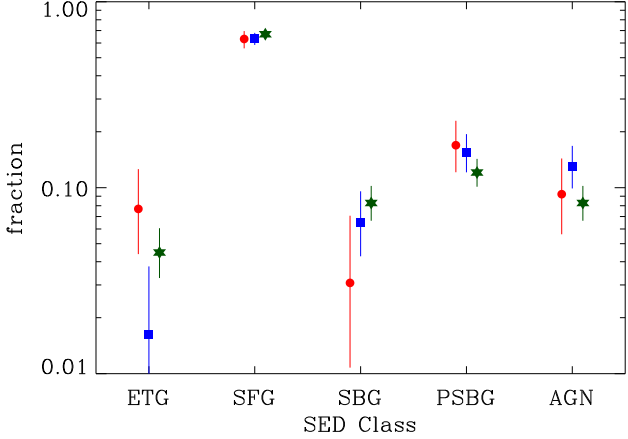}}
\end{minipage}
\end{center}
\caption{The fractions of IR supercluster members ($\rm{z} \cup
  \rm{z}_p$ sample) in different SED classes in the three different
  environments. Different symbols identify the three different
  regions, as in Fig.~\ref{f:3irlf}.}
\label{f:sedregions}
\end{figure}

The similarity between the different sSFR--\mste distributions
suggests that similar modes of star formation take place in galaxies
in different environments.  This similarity has been noted before in
different data sets \citep{Peng+10}.  Additional support for this
result comes from the analysis of the fractions of IR supercluster
members in different SED classes. These fractions are displayed in
Fig.~\ref{f:sedregions} for the different regions of the supercluster
(results are displayed for the $\rm{z} \cup \rm{z}_p$ sample; very
similar results are found for the z sample, and are not shown
here). They are clearly very similar, except perhaps for a very
marginal excess of ETGs in the core region.  The fraction of AGNs
among IR-emitting galaxies is similar to that found in
\citetalias{Edwards-radio} and in other galaxy clusters
\citep[e.g.][]{Geach+09,Krick+09,Chung+10}.

\subsection{Comparison with previous results}
\label{s:comp}
We compare the IR LF of A1763 with those of \citet{Bai+09},
\citet{Tran+09}, and \citet{Chung+10}, for which the parameters of the
best-fit Schechter function are available. Ideally, one would like to
compare IR LFs obtained within regions of similar galaxy number
densities, to highlight differences due to different {\em fractions}
of IR-emitting galaxies. Since previous determinations have been
limited to the inner, virialized cluster regions, we consider in this
comparison only the IR LF of the core region of A1763.

The areas where the LFs of \citet{Bai+09}, \citet{Tran+09},
\citet{Chung+10}, and the A1763 core have been derived correspond to
circular regions of effective limiting radii 0.90, 0.74, 0.82, and
0.65, in units of the respective cluster $\rm{r}_{200}$. We derive the
virial radii of the clusters from their velocity dispersions
\citep[taken from][]{Biviano+96,QRW96,Fisher+98,Barrena+02} via the
relation of \citet[][Appendix A]{MM07}. The effective limiting radii
of the four clusters are similar, but not identical. We therefore
apply scaling factors to the cluster IR LFs proportional to the
estimated number densities of normal galaxies within these limiting
radii.  We compute these projected densities as in Appendix B.2 of
\citet{Mamon+10}, using the individual cluster virial radii and the
model profile of \citet{NFW97} with concentration $c \simeq 3$
\citep[a typical value for rich clusters; see, e.g.][]{BP09}.  Setting
to unity the scaling factor for the IR LF of the A1763 core, the other
scaling factors are 1.40, 1.14, and 1.27, for the LF of
\citet{Bai+09}, \citet{Tran+09}, and \citet{Chung+10}, respectively.

\begin{figure}
\begin{center}
\begin{minipage}{0.5\textwidth}
\resizebox{\hsize}{!}{\includegraphics{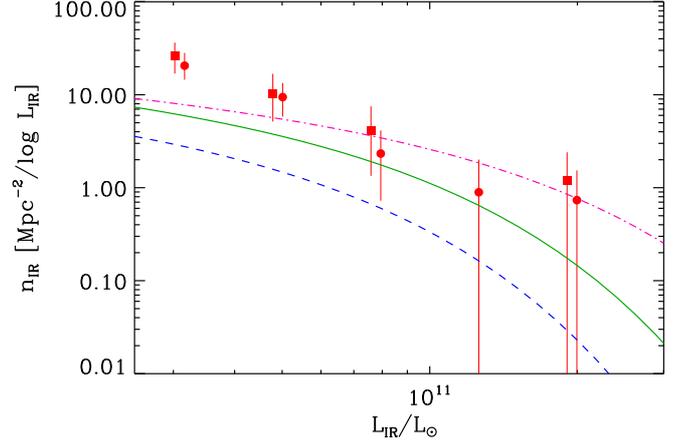}}
\end{minipage}
\end{center}
\caption{The IR LF of the A1763 core region (same as in
  Fig.~\ref{f:3irlf}; filled red dots: $\rm{z} \cup \rm{z}_p$ sample;
  filled red squares -- slightly displaced along the x-axis for
  clarity -- : $\rm{z}$-only sample; 1$\sigma$ error bars), compared
  to the best-fit Schechter IR LF of \citet[][solid green
    curve]{Bai+09}, \citet[][dash-dotted pink curve]{Chung+10}, and
  \citet[][dashed blue curve]{Tran+09}.  Note that the Schechter IR
  LFs have been corrected to take into account the different survey
  areas and the different cluster redshifts, as described in the
  text.}
\label{f:irlflit}
\end{figure}

In addition to the density correction, since the different clusters
are located at different redshifts, we rescale the best-fit Schechter
parameters obtained for these clusters to the redshift of A1763,
adopting the evolution relation of \citet{Bai+09}.  The result is
shown in Fig.~\ref{f:irlflit}. We note that to compare the different
luminosity functions, we divide the number densities of the A1763 core
IR LF by the logarithmic interval we used for the binning, 0.2.

The IR LF of the A1763 core lies significantly above all other IR LFs
at the faint end, while it is consistent with them at the bright end.
It is most similar to the IR LF established by \citet{Chung+10} for
the Bullet cluster. There clearly seems to be a large variance in the
cluster IR LFs, even after correcting for evolutionary effects and
after rescaling for the different galaxy densities in the cluster
areas sampled by the different surveys. Part of the variance is caused
by observational errors, and the IR LF of \citet{Tran+09}, which
appears to lie below that of \citet{Bai+09} in Fig.~\ref{f:irlflit},
is consistent with it within the uncertainties \citep[see Fig.~7
  in][]{Tran+09}. As a source of intrinsic variance, one could
consider the effect of an increasing fraction of IR-emitting galaxies
with clustercentric radius \citep[e.g.][]{Bai+09,Haines+09b}. However,
this trend is far too small to account for the variance we see in the
IR LFs, given that they were obtained within rather similar limiting
effective radii. Moreover, among the four LFs displayed in
Fig.~\ref{f:irlflit}, that of the A1763 core has been determined
within the smallest effective radius, and yet it appears to lie above
all the others.

\begin{figure}
\begin{center}
\begin{minipage}{0.5\textwidth}
\resizebox{\hsize}{!}{\includegraphics{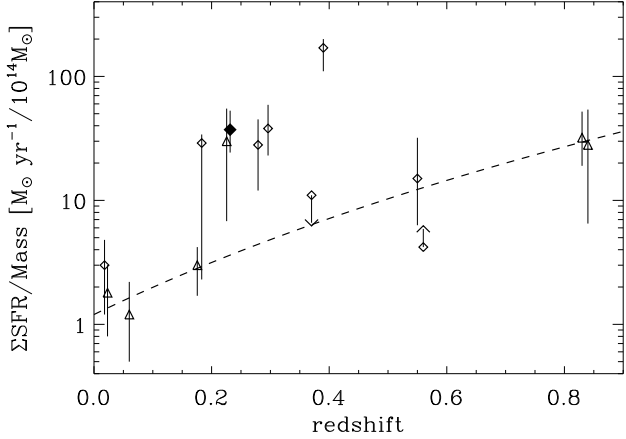}}
\end{minipage}
\end{center}
\caption{$\Sigma$SFR/M$_{200}$ (in units of $\rm{M}_{\odot} \,
  \rm{yr}^{-1}/10^{14} \, \rm{M}_{\odot})$, where $\Sigma$SFR is
  estimated within a region of radius $0.5 \, \rm{r}_{200}$, as a
  function of redshift for A1763 (filled diamond) and for other
  clusters from the literature. Triangles and open diamonds denote
  values taken from \citet{Bai+09} and \citet{Chung+10}, respectively.
  1$\sigma$ error bars are shown. The downward-directed (respectively
  upward-directed) arrow denotes an upper (respectively lower) limit.
  The curve represents the fitting relation of \citet{Bai+09}
  $\Sigma$SFR/mass $= 1.2 (1+\rm{z})^{5.3}$.}
\label{f:tsfrm}
\end{figure}

Another way of comparing IR LFs for clusters of different masses and
at different redshifts is to look at the variation in the total
cluster SFR within a fixed aperture -- in units of $\rm{r}_{200}$ --
normalized by the cluster total mass, $\rm{M}_{200}$
\citep{Geach+06,Bai+09,Chung+10}. The mass of the virial region of
A1763, $\rm{M}_{200}=9.9 \times 10^{14} \, \rm{M}_{\odot}$, is
obtained from the values of $\rm{r}_{200}$ and $\rm{v}_{200}$ (see
Sect.~\ref{s:irlfenv}). The estimate is based on 357 member galaxies
(see Sect.~\ref{s:members}). According to \citet{Biviano+06}, a mass
estimate based on a galaxy sample of this size has an uncertainty of
$\approx 25$\%.

Following \citet{Bai+07,Bai+09}, we evaluate $\Sigma$SFR within
$\rm{R} \leq 0.5 \, \rm{r}_{200}$ and normalize it by the global
cluster mass $\rm{M}_{200}$. The value is reported in
Table~\ref{t:sfr}. The quoted error includes the contribution from
both the $\Sigma$SFR uncertainty (estimated via a bootstrap procedure)
and the mass uncertainty, the latter being the main source of
error. To compare it with the determinations of \citet{Bai+07} we need
to apply a correction to account for the different \lir~ limit of our
IR LF ($2.5 \cdot 10^{10} \, \rm{L}_{\odot}$) and that of
\citet{Bai+07} ($1.2 \cdot 10^{10} \, \rm{L}_{\odot}$).  Using the IR
LF of \citet{Bai+09} evolved to the mean redshift of A1763, we
estimate a correction factor of 1.4. We plot the corrected value for
the A1763 $0.5 \, \rm{r}_{200}$ region in Fig.~\ref{f:tsfrm} together
with the values for other clusters taken from the literature
\citep{Geach+06,Bai+09,Chung+10} and based on {\em IRAS}
\citep{MBK00}, {\em ISO}
\citep{Fadda+00,Duc+02b,Metcalfe+03,Biviano+04,Duc+04,Coia+05a}, and
      {\em Spitzer}
      \citep{Geach+06,Bai+06,Bai+07,Bai+09,Haines+09,Chung+10}
      data. The value for A1763 lies in the same locus of the diagram
      as other clusters at similar redshifts. There is a trend of
      increasing $\Sigma$SFR/mass with redshift, a sort of IR
      Butcher-Oemler effect \citep{BO84,STH08,Haines+09b,TDI09}.  This
      trend has been noticed before and modeled with a power-law
      relation in $(1+\rm{z})$ by \citet{Geach+06} and \citet{Bai+09},
      mimicking the trend found by \citet{CBFC04} for the number of
      ultra-LIRG radio sources in the field, or the trend found by
      \citet{Kodama+04} for the $\Sigma$SFR of cluster galaxies, based
      on optical-line spectroscopic estimates of the galaxy SFRs.  The
      best-fit relation obtained by \citet{Bai+09} is shown in
      Fig.~\ref{f:tsfrm}. It clearly fails to fit the data in the
      $\rm{z}$-range 0.2--0.4. The quantity $\Sigma$SFR/mass appears
      to evolve rapidly from $\rm{z} \approx 0.4$ to 0, while it
      remains almost constant for $\rm{z} \ga 0.4$.

To explore the environmental dependence of the mass-normalized SFR, we
would need to determine the masses of the filament and outskirts
regions, but this is not possible since these regions do not
correspond to virialized, collapsed structures.  With the A1763 mass
$\rm{M}_{200}$ and $r$-band richness within $\rm{r}_{200}$, we define
a mass-to-richness ratio that we use to determine $\Sigma$SFR/richness
at the average redshift of A1763 from the evolutionary relation of
\citet{Bai+09}. This is displayed in Fig.~\ref{f:tsfrr} with a dashed
line and allows an indirect comparison of Fig.~\ref{f:tsfrm} with
Fig.~\ref{f:tsfrr}, where we show the $\Sigma$SFR/richness of the
$\rm{R} \leq 0.5 \, \rm{r}_{200}$ region and also of the core,
filament, and outskirts regions (see Sect.~\ref{s:irlfenv}). We can
summarize the results illustrated in Figs.~\ref{f:tsfrm} and
\ref{f:tsfrr} by saying that the SFR per galaxy increases with z in
clusters, and it is not a monotonic function of the density of the
environment.

\section{Discussion}
\label{s:disc}
Our analysis of the IR LF of the A1763 supercluster has confirmed our
findings of \citetalias{Fadda+08}, namely the filament is the most
probable site of galaxy star formation. We have shown that the IR LF
of filament galaxies lies above those of the core and the outskirts,
when these three IR LFs are normalized by the average densities of
normal, $r$-band selected, galaxies in the three regions
(Fig.~\ref{f:3irlf}, right panels). In other words, filament galaxies
have a higher chance of being IR-emitting than both core and outskirts
galaxies. Since we have corrected the IR LFs for the contribution of
galaxies with AGNs, the excess fraction of IR galaxies in the filament
can also be read as an excess fraction of star-forming
galaxies\footnote{Had we omitted to correct the IR LFs for the
  presence of AGNs, the results of this paper would not have been
  significantly affected.}. The result of our analysis extends the
original finding of \citetalias{Fadda+08} by showing that there is an
excess fraction of star-forming galaxies in the filament relative to
other supercluster regions, at all levels of star-formation down to
our \lir~ completeness limit, which corresponds to $\rm{SFR} \simeq 4
\, M_{\odot} \, \mbox{yr}^{-1}$ (see Sect.~\ref{s:irlfenv}).

Several authors have previously noted the increasing fraction of
IR-emitting, star-forming galaxies with increasing cluster-centric
distance and the lack of LIRGs in cluster cores
\citep{Bai+06,Bai+09,Haines+09b,TDI09,Davies+10,Finn+10}.  Our finding
indicates that the relation is not simply one of galaxy SFR with
cluster-centric distance or local density (see Fig.~\ref{f:tsfrr}).
Galaxy filaments are regions of intermediate galaxy densities between
cluster cores and cluster outskirts, and galaxies in the A1763
filament are not farther out from the A1763 cluster center than
galaxies in the A1763 outskirts (see Fig.~\ref{f:regions}). The higher
fraction of star-forming galaxies in medium-density environments has
already been noted in other IR \citep{Koyama+08,Gallazzi+09,Koyama+10}
or optical \citep{BPB07,PR07,Boue+08,Porter+08,BPBB09} studies of
superclusters.

An interesting aspect of the higher fraction of star-forming galaxies
in the A1763 filament is that these galaxies are relatively massive,
with a relatively low sSFR (see Fig.~\ref{f:regions}). The sSFR-\mste~
relation is very similar in the core, the filament, and the outskirts
regions (see Fig.~\ref{f:ssfrmste}), i.e. at a given \mste~ filament
galaxies do not have enhanced sSFRs.  Both the universality of the
sSFR--\mste~ relation across different environments and the relatively
low-sSFR of dust-reddened high-SFR galaxies in medium-density
environments have been noted before \citep{Peng+10,Gallazzi+09}.

The similarity of the sSFR-\mste~ relation of galaxies in different
supercluster regions suggests that the regions share a similar mode of
star formation.  This conclusion is supported by the similarity of the
SED-class distributions in the different regions
(Fig.~\ref{f:sedregions}). About 65~\% of the IR-emitting galaxies in
all the supercluster regions belong to the normal SFG SED-class.  Only
$\sim 6$~\% and, respectively, $\sim 15$~\% belong to the SBG and,
respectively, PSBG classes, $\sim 10$~\% to the AGN class, and $\sim
4$~\% (slightly more in the core) to the ETG class. These fractions
indicate that, overall, the dominant mode of star formation in
IR-emitting galaxies across the whole supercluster is that of normal
late-type galaxies. \citet{Haines+11b} reached the same conclusion
from their analysis of IR and UV data for the nearby Shapley
supercluster.

Even if the SFG SED class characterizes most of the IR-emitting
galaxies, the brightest of them, LIRGs, mostly belong to the SBG SED
class (10 out of 18 LIRGs belong to the SBG SED class; see also
Fig.~\ref{f:fractions}). It is known that LIRGs are mostly powered by
starbursts \citep[e.g.][]{daCunha+10,Fadda+10,Finn+10}, which occur as
a consequence of close galaxy-galaxy interactions
\citep[e.g.][]{SB51,NW83,Bushouse87,Sanders+88,Barnes92,Hwang+10,TCB10}.
In A1763, LIRGs are more frequently found in the filament than in
other regions of the supercluster (Fig.~\ref{f:regions}).  In cluster
cores, interactions are frequent but do not significantly affect the
interacting galaxies because of the very high speed of these
encounters \citep[repeated fast encounters might however be relevant
  for dwarf galaxies,][]{Moore+96}. In filaments, on the other hand,
the frequency of galaxy-galaxy interactions is still relatively high,
and they occur at a relatively low speed, since the filament
environment is characterized by higher galaxy densities than the
field, and lower velocity dispersions than the cluster core
\citepalias{Fadda+08}.  The tidal \citep[or resonance, see][]{DVFH09}
effects of galaxy-galaxy interactions are stronger when the collisions
occur at small relative velocities \citep{Mamon96,MH97}, so filaments
(and poor groups) are the ideal environment for significant effects to
result from galaxy-galaxy interactions. These interactions may
sporadically lead to (major) mergers.

SBGs are likely to evolve into galaxies of the PSBG class. Further
evolution is hard to predict.  It has been argued that AGNs are a late
outcome of the starburst process
\citep[e.g.][]{UFM99,Emonts+06,Younger+09}. In this case, we could
estimate that starburst episodes are affecting, or have affected in
the past $\sim 1$ Gyr, $\sim 30$\% of all the IR-emitting galaxies in
the A1763.  According to \citet{Wild+09}, there is a similar fraction
of red sequence galaxies that evolved through a starburst. Filament
SBGs and their outcomes are therefore an important path of galaxy
evolution in and around clusters, even if they do not represent the
dominant channel to move galaxies from the blue cloud to the red
sequence, since the observed fraction of PSBGs in clusters is too
small \citep{DeLucia+09}.

The relative numbers of SBGs and PSBGs probably reflects the relative
durations of the starburst and post-starburst phases, that is $\sim
1/3$--1/2. If the post-starburst phase typically lasts $\sim 1$~Gyr
\citep{Hogg+06,Goto07}, this ratio implies that the starburst phase
lasts $\sim 0.4$~Gyr, close to recent estimates \citep{McQuinn+10}. As
a consequence, the SBG infall speed ($\sim 1$ Mpc/Gyr in projection)
is insufficient for them to travel along the whole filament into the
cluster before the starburst phase is over, so they enter the cluster
as PSBGs (or, maybe, AGNs).  We do however observe SBGs in cluster
cores.  How do they originate?  Part of them are likely to be found in
the center only because of projection effects. Part of them may form
in subclusters as they are tidally compressed by the cluster
gravitational field \citep{Oemler+09}.  Since the accretion of groups
(i.e. subclusters) in clusters increases with $\rm{z}$
\citep{ELYC01,vandenBosch02}, higher-$\rm{z}$ clusters are expected to
display a higher fraction of SBGs in their central regions, as indeed
found by \citet{Dressler+09}.

As far as the evolution of the IR LF is concerned, our results confirm
the results of \citet{Bai+09}, namely that the number density of
IR-emitting galaxies in clusters increases with $\rm{z}$ at all \lir.
Similarly, the total SFR of cluster galaxies per cluster mass also
increases with $\rm{z}$ \citep{Geach+06,Bai+09}, at least until
$\rm{z} \approx 0.4$ (see Fig.~\ref{f:tsfrm}).  As suggested by
\citet{Finn+10}, this evolution is likely to result from the
combination of a general decline in the SFR of field galaxies
(consequence of the gradual exhaustion of their gas reservoirs)
coupled to a decrease in the infall rate of field galaxies into
clusters \citep{ELYC01,vandenBosch02} and to a quenching process at
work in the cluster environment, presumably ram-pressure stripping
\citep{GG72,QMB00}.

This evolution appears to accelerate at $\rm{z} \la 0.4$ as expected
if it is linked to the accretion rate of field galaxies, which peaks
at relatively low-$\rm{z}$ for cluster-sized halos
\citep{vandenBosch02}.  More data are needed to confirm that the
evolution is indeed accelerated at $\rm{z} \la 0.4$. The current
sample of clusters on which the relation of \citet{Bai+09} is based is
not complete, and we cannot exclude that the clusters
that show excess star formation at $0.2 \la z \la 0.4$ may be a biased
set. If all these clusters are currently undergoing mergers, their
excess of star formation may be interpreted as the result of contaminating
the pristine cluster galaxy population with the presumably younger
galaxy population of infalling groups \citep{Chung+10}. A detailed
dynamical analysis of the A1763 cluster will be the subject of a
forthcoming paper in this series, but indications that this cluster is
far from relaxation have already been provided in
\citetalias{Fadda+08}.

In summary, the evolution of the number density of IR-emitting
galaxies in cluster cores could result from the competing processes of
accretion of star-forming field galaxies, and quenching. It is
possible that most IR-emitting galaxies in cluster cores are
star-forming galaxies recently infallen from the field that have not
yet spent sufficient time in the cluster environment for their
star-formation to be quenched. The radially elongated orbits of star
forming galaxies in clusters is also suggestive of their recent infall
\citep{BK04}. If the quenching process is fast enough, one expects to
see an environmental dependence of the fraction of IR-emitting
galaxies but not of their intrinsic properties. This is what is
indicated by the similarities of the sSFR-\mste~ relations (see
Fig.~\ref{f:ssfrmste}), of the distributions of galaxies in SED
classes (see Fig.~\ref{f:sedregions}), and of the shape of the IR LFs
\citep[see Table~\ref{t:fits}; see also the results of][for the
  Shapley supercluster]{Haines+11a} across the different supercluster
regions.

\section{Summary and conclusions}
\label{s:summ}
We determine the IR LF of the A1763 supercluster of star-forming
galaxies at $\rm{z} \simeq 0.23$. Supercluster members are selected in
a sample of 24-\mic-detected sources on the basis of their
spectroscopic and photometric redshifts.  Total \lir~ and \mste~ for
supercluster members are obtained by fitting their SEDs. AGNs are
  identified by their SEDs and other methods
  \citepalias[see][]{Edwards-radio} and their contribution removed
  from the IR LFs.  Comparison with \lir-estimates obtained from
monochromatic 24~\mic~ luminosities shows that a good photometric
coverage of the galaxy SEDs is important for accurate \lir-estimates.

We show that the IR LF changes according to the supercluster
environment. We define three environments: the cluster core, the
large-scale filament, and the cluster outskirts, in order of
decreasing galaxy density. By normalizing the IR LFs with the average
number densities of optically-selected galaxies, we show that the
filament hosts the highest fraction of IR-emitting galaxies at all \lir.
Similarly, the filament region contains the highest total SFR per unit
galaxy. At the other extreme lies the core region,
where LIRGs are almost absent. The IR LF of the cluster outskirts
(excluding the filament region) is intermediate between those of the
filament and the core.

We do not find any environmental dependence of the \mbox{sSFR-\mste}~
relation.  Most high-star forming galaxies in the supercluster are
also massive, and the excess population of LIRGs in the filament
region is due to massive galaxies with normal sSFRs for their \mste,
that is to say, relatively low sSFRs.

Galaxies of the different regions have very similar fractions of
SED-classes. Normal, SFGs dominate; SBGs dominate at the bright
  end of the IR LF; AGNs contribute only $\sim 10$\% in
fraction.

Comparison with previous results from the literature confirms the
evolution of cluster IR LF found by \cite{Bai+09}, as well as the
evolution of total cluster SFR divided by cluster mass
\citep{Geach+06,Bai+09,Chung+10}. The evolution is faster
at $\rm{z} \la 0.4$ than at higher $\rm{z}$, unless the clusters that
have so far been investigated in the IR at $0.2 \la z \la 0.4$ are a
biased set of dynamically young systems, in which the presence of
infalling groups biases the estimates of total cluster SFR high.

We discuss these results by drawing a scenario for the evolution of
galaxies in and around clusters. Massive star-forming galaxies exist
in medium-density environments at $\rm{z} \sim 0.2$; about two-thirds
of them have a mode of star formation resembling that of normal
late-type galaxies. As these galaxies enter the cluster environment,
they suffer ram-pressure stripping and evolve into passive
galaxies. The remaining fraction is undergoing or has recently
experienced starbursts, probably induced by galaxy-galaxy interactions
(or mergers). They enter the cluster as PSBGs. Together, these two
paths of galaxy evolution lead to the build-up of the red sequence in
clusters.

In future papers of this series, we will present the spectroscopic
catalog of the A1763 region and the new UV data we have obtained from
{\em GALEX} observations; we will investigate the dynamics of the
A1763 cluster and the spectral properties of the galaxies in the A1763
supercluster. We also plan to determine morphologies for A1763
supercluster galaxies, and to deepen our investigation into the
low-\lir~ regime with {\em Herschel} satellite observations.

\begin{acknowledgements}
  We warmly thank the anonymous referee for the careful reports that
  have helped us to significantly improve this paper.  We acknowledge
  useful discussions with Sun Mi Chung, Nicholas Lee, Claudia
  Maraston, Paola Popesso, George Rieke, and Laura Silva. AB and
  FD acknowledge the hospitality of IPAC at Caltech, and AB also the
  hospitality of the Institut d'Astrophysique de Paris, during the
  preparation of this work.

  Partial financial support for this research has been provided by the
  Agenzia Spaziale Italiana through the projects ``IR studies of
  clusters of galaxies from $\rm{z}=3$ to $\rm{z}=0$'', and ``Star
  formation in galaxy superclusters with GALEX'', and by NASA through
  an award issued by JPL/Caltech.

  This research has made use of NASA's Astrophysics Data System, of
  NED, which is operated by JPL/Caltech, under contract with NASA, and
  of SDSS, which has been funded by the Sloan Foundation, NSF, the US
  Department of Energy, NASA, the Japanese Monbukagakusho, the Max
  Planck Society, and the Higher Education Funding Council of England.
  The SDSS is managed by the participating institutions
  (www.sdss.org/collaboration/credits.html).

\end{acknowledgements}

\bibliography{master}

\begin{thebibliography}{124}
\expandafter\ifx\csname natexlab\endcsname\relax\def\natexlab#1{#1}\fi

\bibitem[{{Babbedge} {et~al.}(2006){Babbedge}, {Rowan-Robinson}, {Vaccari},
  {Surace}, {Lonsdale}, {Clements}, {Fang}, {Farrah}, {Franceschini},
  {Gonzalez-Solares}, {Hatziminaoglou}, {Lacey}, {Oliver}, {Onyett},
  {P{\'e}rez-Fournon}, {Polletta}, {Pozzi}, {Rodighiero}, {Shupe}, {Siana}, \&
  {Smith}}]{Babbedge+06}
{Babbedge}, T.~S.~R., {Rowan-Robinson}, M., {Vaccari}, M., {et~al.} 2006,
  \mnras, 370, 1159

\bibitem[{{Bai} {et~al.}(2007){Bai}, {Marcillac}, {Rieke}, {Rieke}, {Tran},
  {Hinz}, {Rudnick}, {Kelly}, \& {Blaylock}}]{Bai+07}
{Bai}, L., {Marcillac}, D., {Rieke}, G.~H., {et~al.} 2007, \apj, 664, 181

\bibitem[{{Bai} {et~al.}(2009){Bai}, {Rieke}, {Rieke}, {Christlein}, \&
  {Zabludoff}}]{Bai+09}
{Bai}, L., {Rieke}, G.~H., {Rieke}, M.~J., {Christlein}, D., \& {Zabludoff},
  A.~I. 2009, \apj, 693, 1840

\bibitem[{{Bai} {et~al.}(2006){Bai}, {Rieke}, {Rieke}, {Hinz}, {Kelly}, \&
  {Blaylock}}]{Bai+06}
{Bai}, L., {Rieke}, G.~H., {Rieke}, M.~J., {et~al.} 2006, \apj, 639, 827

\bibitem[{{Barnes}(1992)}]{Barnes92}
{Barnes}, J.~E. 1992, \apj, 393, 484

\bibitem[{{Barrena} {et~al.}(2002){Barrena}, {Biviano}, {Ramella}, {Falco}, \&
  {Seitz}}]{Barrena+02}
{Barrena}, R., {Biviano}, A., {Ramella}, M., {Falco}, E.~E., \& {Seitz}, S.
  2002, \aap, 386, 816

\bibitem[{{Beers} {et~al.}(1990){Beers}, {Flynn}, \& {Gebhardt}}]{BFG90}
{Beers}, T.~C., {Flynn}, K., \& {Gebhardt}, K. 1990, \aj, 100, 32

\bibitem[{{Bell} {et~al.}(2003){Bell}, {McIntosh}, {Katz}, \&
  {Weinberg}}]{Bell+03}
{Bell}, E.~F., {McIntosh}, D.~H., {Katz}, N., \& {Weinberg}, M.~D. 2003, \apjs,
  149, 289

\bibitem[{{Bernardi} {et~al.}(2010){Bernardi}, {Shankar}, {Hyde}, {Mei},
  {Marulli}, \& {Sheth}}]{Bernardi+10}
{Bernardi}, M., {Shankar}, F., {Hyde}, J.~B., {et~al.} 2010, \mnras, 404, 2087

\bibitem[{{Biviano}(2008)}]{Biviano08}
{Biviano}, A. 2008, arXiv:0811.3535

\bibitem[{{Biviano} {et~al.}(1996){Biviano}, {Durret}, {Gerbal}, {Le Fevre},
  {Lobo}, {Mazure}, \& {Slezak}}]{Biviano+96}
{Biviano}, A., {Durret}, F., {Gerbal}, D., {et~al.} 1996, \aap, 311, 95

\bibitem[{{Biviano} \& {Katgert}(2004)}]{BK04}
{Biviano}, A. \& {Katgert}, P. 2004, \aap, 424, 779

\bibitem[{{Biviano} {et~al.}(2004){Biviano}, {Metcalfe}, {McBreen}, {Altieri},
  {Coia}, {Kessler}, {Kneib}, {Leech}, {Okumura}, {Ott}, {Perez-Martinez},
  {Sanchez-Fernandez}, \& {Schulz}}]{Biviano+04}
{Biviano}, A., {Metcalfe}, L., {McBreen}, B., {et~al.} 2004, \aap, 425, 33

\bibitem[{{Biviano} {et~al.}(2006){Biviano}, {Murante}, {Borgani}, {Diaferio},
  {Dolag}, \& {Girardi}}]{Biviano+06}
{Biviano}, A., {Murante}, G., {Borgani}, S., {et~al.} 2006, \aap, 456, 23

\bibitem[{{Biviano} \& {Poggianti}(2009)}]{BP09}
{Biviano}, A. \& {Poggianti}, B.~M. 2009, \aap, 501, 419

\bibitem[{{Bothwell} {et~al.}(2011){Bothwell}, {Kennicutt}, {Johnson}, {Wu},
  {Lee}, {Dale}, {Engelbracht}, {Calzetti}, \& {Skillman}}]{Bothwell+11}
{Bothwell}, M.~S., {Kennicutt}, R.~C., {Johnson}, B.~D., {et~al.} 2011,
  arXiv:1104.0929

\bibitem[{{Bou{\'e}} {et~al.}(2008){Bou{\'e}}, {Durret}, {Adami}, {Mamon},
  {Ilbert}, \& {Cayatte}}]{Boue+08}
{Bou{\'e}}, G., {Durret}, F., {Adami}, C., {et~al.} 2008, \aap, 489, 11

\bibitem[{{Braglia} {et~al.}(2007){Braglia}, {Pierini}, \&
  {B{\"o}hringer}}]{BPB07}
{Braglia}, F., {Pierini}, D., \& {B{\"o}hringer}, H. 2007, \aap, 470, 425

\bibitem[{{Braglia} {et~al.}(2009){Braglia}, {Pierini}, {Biviano}, \&
  {B{\"o}hringer}}]{BPBB09}
{Braglia}, F.~G., {Pierini}, D., {Biviano}, A., \& {B{\"o}hringer}, H. 2009,
  \aap, 500, 947

\bibitem[{{Brammer} {et~al.}(2008){Brammer}, {van Dokkum}, \& {Coppi}}]{BvDC08}
{Brammer}, G.~B., {van Dokkum}, P.~G., \& {Coppi}, P. 2008, \apj, 686, 1503

\bibitem[{{Bushouse}(1987)}]{Bushouse87}
{Bushouse}, H.~A. 1987, \apj, 320, 49

\bibitem[{{Butcher} \& {Oemler}(1984)}]{BO84}
{Butcher}, H. \& {Oemler}, Jr., A. 1984, \apj, 285, 426

\bibitem[{{Calzetti} {et~al.}(2000){Calzetti}, {Armus}, {Bohlin}, {Kinney},
  {Koornneef}, \& {Storchi-Bergmann}}]{Calzetti+00}
{Calzetti}, D., {Armus}, L., {Bohlin}, R.~C., {et~al.} 2000, \apj, 533, 682

\bibitem[{{Cavagnolo} {et~al.}(2009){Cavagnolo}, {Donahue}, {Voit}, \&
  {Sun}}]{Cavagnolo+09}
{Cavagnolo}, K.~W., {Donahue}, M., {Voit}, G.~M., \& {Sun}, M. 2009, \apjs,
  182, 12

\bibitem[{{Chung} {et~al.}(2010){Chung}, {Gonzalez}, {Clowe}, {Markevitch}, \&
  {Zaritsky}}]{Chung+10}
{Chung}, S.~M., {Gonzalez}, A.~H., {Clowe}, D., {Markevitch}, M., \&
  {Zaritsky}, D. 2010, \apj, 725, 1536

\bibitem[{{Coia} {et~al.}(2005{\natexlab{a}}){Coia}, {McBreen}, {Metcalfe},
  {Biviano}, {Altieri}, {Ott}, {Fort}, {Kneib}, {Mellier},
  {Miville-Desch{\^e}nes}, {O'Halloran}, \& {Sanchez-Fernandez}}]{Coia+05b}
{Coia}, D., {McBreen}, B., {Metcalfe}, L., {et~al.} 2005{\natexlab{a}}, \aap,
  431, 433

\bibitem[{{Coia} {et~al.}(2005{\natexlab{b}}){Coia}, {Metcalfe}, {McBreen},
  {Biviano}, {Smail}, {Altieri}, {Kneib}, {McBreen}, {Sanchez-Fernandez}, \&
  {O'Halloran}}]{Coia+05a}
{Coia}, D., {Metcalfe}, L., {McBreen}, B., {et~al.} 2005{\natexlab{b}}, \aap,
  430, 59

\bibitem[{{Collister} \& {Lahav}(2004)}]{CL04}
{Collister}, A.~A. \& {Lahav}, O. 2004, \pasp, 116, 345

\bibitem[{{Cowie} {et~al.}(2004){Cowie}, {Barger}, {Fomalont}, \&
  {Capak}}]{CBFC04}
{Cowie}, L.~L., {Barger}, A.~J., {Fomalont}, E.~B., \& {Capak}, P. 2004, \apjl,
  603, L69

\bibitem[{{da Cunha} {et~al.}(2010){da Cunha}, {Eminian}, {Charlot}, \&
  {Blaizot}}]{daCunha+10}
{da Cunha}, E., {Eminian}, C., {Charlot}, S., \& {Blaizot}, J. 2010, \mnras,
  403, 1894

\bibitem[{{Davies} {et~al.}(2010){Davies}, {Baes}, {Bendo}, {Bianchi},
  {Bomans}, {Boselli}, {Clemens}, {Corbelli}, {Cortese}, {Dariush}, {de Looze},
  {di Serego Alighieri}, {Fadda}, {Fritz}, {Garcia-Appadoo}, {Gavazzi},
  {Giovanardi}, {Grossi}, {Hughes}, {Hunt}, {Jones}, {Madden}, {Pierini},
  {Pohlen}, {Sabatini}, {Smith}, {Verstappen}, {Vlahakis}, {Xilouris}, \&
  {Zibetti}}]{Davies+10}
{Davies}, J.~I., {Baes}, M., {Bendo}, G.~J., {et~al.} 2010, \aap, 518, L48

\bibitem[{{De Lucia} {et~al.}(2009){De Lucia}, {Poggianti}, {Halliday},
  {Milvang-Jensen}, {Noll}, {Smail}, \& {Zaritsky}}]{DeLucia+09}
{De Lucia}, G., {Poggianti}, B.~M., {Halliday}, C., {et~al.} 2009, \mnras, 400,
  68

\bibitem[{{DeGroot}(1987)}]{DeGroot87}
{DeGroot}, M. 1987, Probability and Statistics, Second Edition (Addison-Wesley
  Publishing Co., Reading, MA)

\bibitem[{{den Hartog} \& {Katgert}(1996)}]{dHK96}
{den Hartog}, R. \& {Katgert}, P. 1996, \mnras, 279, 349

\bibitem[{{D'Onghia} {et~al.}(2010){D'Onghia}, {Vogelsberger},
  {Faucher-Giguere}, \& {Hernquist}}]{DVFH09}
{D'Onghia}, E., {Vogelsberger}, M., {Faucher-Giguere}, C., \& {Hernquist}, L.
  2010, \apj, 725, 353

\bibitem[{{Dressler} {et~al.}(2009){Dressler}, {Rigby}, {Oemler}, {Fritz},
  {Poggianti}, {Rieke}, \& {Bai}}]{Dressler+09}
{Dressler}, A., {Rigby}, J., {Oemler}, A., {et~al.} 2009, \apj, 693, 140

\bibitem[{{Duc} {et~al.}(2004){Duc}, {Fadda}, {Poggianti}, {Elbaz}, {Biviano},
  {Flores}, {Moorwood}, {Franceschini}, \& {Cesarsky}}]{Duc+04}
{Duc}, P.-A., {Fadda}, D., {Poggianti}, B., {et~al.} 2004, in IAU Colloq. 195:
  Outskirts of Galaxy Clusters: Intense Life in the Suburbs, ed. A.~{Diaferio},
  347--351

\bibitem[{{Duc} {et~al.}(2002){Duc}, {Poggianti}, {Fadda}, {Elbaz}, {Flores},
  {Chanial}, {Franceschini}, {Moorwood}, \& {Cesarsky}}]{Duc+02b}
{Duc}, P.-A., {Poggianti}, B.~M., {Fadda}, D., {et~al.} 2002, \aap, 382, 60

\bibitem[{{Edwards} {et~al.}(2010{\natexlab{a}}){Edwards}, {Fadda}, {Biviano},
  \& {Marleau}}]{Edwards+10}
{Edwards}, L.~O.~V., {Fadda}, D., {Biviano}, A., \& {Marleau}, F.~R.
  2010{\natexlab{a}}, \aj, 139, 434 (Paper 1)

\bibitem[{{Edwards} {et~al.}(2010{\natexlab{b}}){Edwards}, {Fadda}, {Frayer},
  {Lima Neto}, \& {Durret}}]{Edwards-radio}
{Edwards}, L.~O.~V., {Fadda}, D., {Frayer}, D.~T., {Lima Neto}, G.~B., \&
  {Durret}, F. 2010{\natexlab{b}}, \aj, 140, 1891 (Paper 2)

\bibitem[{{Efron} \& {Tibshirani}(1986)}]{ET86}
{Efron}, B. \& {Tibshirani}, R. 1986, Stat. Sci., 1, 54

\bibitem[{{Ellingson} {et~al.}(2001){Ellingson}, {Lin}, {Yee}, \&
  {Carlberg}}]{ELYC01}
{Ellingson}, E., {Lin}, H., {Yee}, H.~K.~C., \& {Carlberg}, R.~G. 2001, \apj,
  547, 609

\bibitem[{{Emonts} {et~al.}(2006){Emonts}, {Morganti}, {Tadhunter}, {Holt},
  {Oosterloo}, {van der Hulst}, \& {Wills}}]{Emonts+06}
{Emonts}, B.~H.~C., {Morganti}, R., {Tadhunter}, C.~N., {et~al.} 2006, \aap,
  454, 125

\bibitem[{{Fadda} {et~al.}(2008){Fadda}, {Biviano}, {Marleau},
  {Storrie-Lombardi}, \& {Durret}}]{Fadda+08}
{Fadda}, D., {Biviano}, A., {Marleau}, F.~R., {Storrie-Lombardi}, L.~J., \&
  {Durret}, F. 2008, \apjl, 672, L9 (Paper 0)

\bibitem[{{Fadda} {et~al.}(2000){Fadda}, {Elbaz}, {Duc}, {Flores},
  {Franceschini}, {Cesarsky}, \& {Moorwood}}]{Fadda+00}
{Fadda}, D., {Elbaz}, D., {Duc}, P.-A., {et~al.} 2000, \aap, 361, 827

\bibitem[{{Fadda} {et~al.}(1996){Fadda}, {Girardi}, {Giuricin}, {Mardirossian},
  \& {Mezzetti}}]{Fadda+96}
{Fadda}, D., {Girardi}, M., {Giuricin}, G., {Mardirossian}, F., \& {Mezzetti},
  M. 1996, \apj, 473, 670

\bibitem[{{Fadda} {et~al.}(2006){Fadda}, {Marleau}, {Storrie-Lombardi},
  {Makovoz}, {Frayer}, {Appleton}, {Armus}, {Chapman}, {Choi}, {Fang},
  {Heinrichsen}, {Helou}, {Im}, {Lacy}, {Shupe}, {Soifer}, {Squires}, {Surace},
  {Teplitz}, {Wilson}, \& {Yan}}]{Fadda+06}
{Fadda}, D., {Marleau}, F.~R., {Storrie-Lombardi}, L.~J., {et~al.} 2006, \aj,
  131, 2859

\bibitem[{{Fadda} {et~al.}(2010){Fadda}, {Yan}, {Lagache}, {Sajina}, {Lutz},
  {Wuyts}, {Frayer}, {Marcillac}, {Le Floc'h}, {Caputi}, {Spoon}, {Veilleux},
  {Blain}, \& {Helou}}]{Fadda+10}
{Fadda}, D., {Yan}, L., {Lagache}, G., {et~al.} 2010, \apj, 719, 425

\bibitem[{{Fasano} \& {Franceschini}(1987)}]{FF87}
{Fasano}, G. \& {Franceschini}, A. 1987, \mnras, 225, 155

\bibitem[{{Finn} {et~al.}(2010){Finn}, {Desai}, {Rudnick}, {Poggianti}, {Bell},
  {Hinz}, {Jablonka}, {Milvang-Jensen}, {Moustakas}, {Rines}, \&
  {Zaritsky}}]{Finn+10}
{Finn}, R.~A., {Desai}, V., {Rudnick}, G., {et~al.} 2010, \apj, 720, 87

\bibitem[{{Fisher} {et~al.}(1998){Fisher}, {Fabricant}, {Franx}, \& {van
  Dokkum}}]{Fisher+98}
{Fisher}, D., {Fabricant}, D., {Franx}, M., \& {van Dokkum}, P. 1998, \apj,
  498, 195

\bibitem[{{Fontana} {et~al.}(2004){Fontana}, {Pozzetti}, {Donnarumma},
  {Renzini}, {Cimatti}, {Zamorani}, {Menci}, {Daddi}, {Giallongo}, {Mignoli},
  {Perna}, {Salimbeni}, {Saracco}, {Broadhurst}, {Cristiani}, {D'Odorico}, \&
  {Gilmozzi}}]{Fontana+04}
{Fontana}, A., {Pozzetti}, L., {Donnarumma}, I., {et~al.} 2004, \aap, 424, 23

\bibitem[{{Gallazzi} {et~al.}(2009){Gallazzi}, {Bell}, {Wolf}, {Gray},
  {Papovich}, {Barden}, {Peng}, {Meisenheimer}, {Heymans}, {van Kampen},
  {Gilmour}, {Balogh}, {McIntosh}, {Bacon}, {Barazza}, {B{\"o}hm}, {Caldwell},
  {H{\"a}u{\ss}ler}, {Jahnke}, {Jogee}, {Lane}, {Robaina}, {Sanchez}, {Taylor},
  {Wisotzki}, \& {Zheng}}]{Gallazzi+09}
{Gallazzi}, A., {Bell}, E.~F., {Wolf}, C., {et~al.} 2009, \apj, 690, 1883

\bibitem[{{Gao} {et~al.}(2008){Gao}, {Navarro}, {Cole}, {Frenk}, {White},
  {Springel}, {Jenkins}, \& {Neto}}]{Gao+08}
{Gao}, L., {Navarro}, J.~F., {Cole}, S., {et~al.} 2008, \mnras, 387, 536

\bibitem[{{Gavazzi}(2009)}]{Gavazzi09}
{Gavazzi}, G. 2009, in Revista Mexicana de Astronomia y Astrofisica Conference
  Series, Vol.~37, 72--78

\bibitem[{{Geach} {et~al.}(2006){Geach}, {Smail}, {Ellis}, {Moran}, {Smith},
  {Treu}, {Kneib}, {Edge}, \& {Kodama}}]{Geach+06}
{Geach}, J.~E., {Smail}, I., {Ellis}, R.~S., {et~al.} 2006, \apj, 649, 661

\bibitem[{{Geach} {et~al.}(2009){Geach}, {Smail}, {Moran}, {Treu}, \&
  {Ellis}}]{Geach+09}
{Geach}, J.~E., {Smail}, I., {Moran}, S.~M., {Treu}, T., \& {Ellis}, R.~S.
  2009, \apj, 691, 783

\bibitem[{{Goto}(2007)}]{Goto07}
{Goto}, T. 2007, \mnras, 381, 187

\bibitem[{{Goto} {et~al.}(2011){Goto}, {Arnouts}, {Malkan}, {Takagi}, {Inami},
  {Pearson}, {Wada}, {Matsuhara}, {Yamauchi}, {Takeuchi}, {Nakagawa}, {Oyabu},
  {Ishihara}, {Sanders}, {Le Floc'h}, {Lee}, {Jeong}, {Serjeant}, \&
  {Sedgwick}}]{Goto+11}
{Goto}, T., {Arnouts}, S., {Malkan}, M., {et~al.} 2011, \mnras, 414, 1903

\bibitem[{{Gunn} \& {Gott}(1972)}]{GG72}
{Gunn}, J.~E. \& {Gott}, J.~R. 1972, \apj, 176, 1

\bibitem[{{Haines} {et~al.}(2011{\natexlab{a}}){Haines}, {Busarello},
  {Merluzzi}, {Smith}, {Raychaudhury}, {Mercurio}, \& {Smith}}]{Haines+11a}
{Haines}, C.~P., {Busarello}, G., {Merluzzi}, P., {et~al.} 2011{\natexlab{a}},
  \mnras, 412, 127

\bibitem[{{Haines} {et~al.}(2011{\natexlab{b}}){Haines}, {Busarello},
  {Merluzzi}, {Smith}, {Raychaudhury}, {Mercurio}, \& {Smith}}]{Haines+11b}
{Haines}, C.~P., {Busarello}, G., {Merluzzi}, P., {et~al.} 2011{\natexlab{b}},
  \mnras, 412, 145

\bibitem[{{Haines} {et~al.}(2009{\natexlab{a}}){Haines}, {Smith}, {Egami},
  {Ellis}, {Moran}, {Sanderson}, {Merluzzi}, {Busarello}, \&
  {Smith}}]{Haines+09b}
{Haines}, C.~P., {Smith}, G.~P., {Egami}, E., {et~al.} 2009{\natexlab{a}},
  \apj, 704, 126

\bibitem[{{Haines} {et~al.}(2009{\natexlab{b}}){Haines}, {Smith}, {Egami},
  {Okabe}, {Takada}, {Ellis}, {Moran}, \& {Umetsu}}]{Haines+09}
{Haines}, C.~P., {Smith}, G.~P., {Egami}, E., {et~al.} 2009{\natexlab{b}},
  \mnras, 396, 1297

\bibitem[{{Haines} {et~al.}(2010){Haines}, {Smith}, {Pereira}, {Egami},
  {Moran}, {Hardegree-Ullman}, {Rawle}, \& {Rex}}]{Haines+10}
{Haines}, C.~P., {Smith}, G.~P., {Pereira}, M.~J., {et~al.} 2010, \aap, 518,
  L19

\bibitem[{{Harrison} \& {Noonan}(1979)}]{HN79}
{Harrison}, E.~R. \& {Noonan}, T.~W. 1979, \apj, 232, 18

\bibitem[{{Hickox} {et~al.}(2009){Hickox}, {Jones}, {Forman}, {Murray},
  {Kochanek}, {Eisenstein}, {Jannuzi}, {Dey}, {Brown}, {Stern}, {Eisenhardt},
  {Gorjian}, {Brodwin}, {Narayan}, {Cool}, {Kenter}, {Caldwell}, \&
  {Anderson}}]{Hickox+09}
{Hickox}, R.~C., {Jones}, C., {Forman}, W.~R., {et~al.} 2009, \apj, 696, 891

\bibitem[{{Hogg} {et~al.}(2006){Hogg}, {Masjedi}, {Berlind}, {Blanton},
  {Quintero}, \& {Brinkmann}}]{Hogg+06}
{Hogg}, D.~W., {Masjedi}, M., {Berlind}, A.~A., {et~al.} 2006, \apj, 650, 763

\bibitem[{{Hwang} {et~al.}(2010){Hwang}, {Elbaz}, {Lee}, {Jeong}, {Park},
  {Lee}, \& {Lee}}]{Hwang+10}
{Hwang}, H.~S., {Elbaz}, D., {Lee}, J.~C., {et~al.} 2010, \aap, 522, A33+

\bibitem[{{Katgert} {et~al.}(2004){Katgert}, {Biviano}, \& {Mazure}}]{KBM04}
{Katgert}, P., {Biviano}, A., \& {Mazure}, A. 2004, \apj, 600, 657

\bibitem[{{Kennicutt}(1998)}]{ken98}
{Kennicutt}, Jr., R.~C. 1998, ARA\&A, 36, 189

\bibitem[{{Knobel} {et~al.}(2009){Knobel}, {Lilly}, {Iovino}, {Porciani},
  {Kova{\v c}}, {Cucciati}, {Finoguenov}, {Kitzbichler}, {Carollo}, {Contini},
  {Kneib}, {LeF{\`e}vre}, {Mainieri}, {Renzini}, {Scodeggio}, {Zamorani},
  {Bardelli}, {Bolzonella}, {Bongiorno}, {Caputi}, {Coppa}, {de la Torre}, {de
  Ravel}, {Franzetti}, {Garilli}, {Kampczyk}, {Lamareille}, {Le Borgne}, {Le
  Brun}, {Maier}, {Mignoli}, {Pello}, {Peng}, {Montero}, {Ricciardelli},
  {Silverman}, {Tanaka}, {Tasca}, {Tresse}, {Vergani}, {Zucca}, {Abbas},
  {Bottini}, {Cappi}, {Cassata}, {Cimatti}, {Fumana}, {Guzzo}, {Koekemoer},
  {Leauthaud}, {Maccagni}, {Marinoni}, {McCracken}, {Memeo}, {Meneux}, {Oesch},
  {Pozzetti}, \& {Scaramella}}]{Knobel+09}
{Knobel}, C., {Lilly}, S.~J., {Iovino}, A., {et~al.} 2009, \apj, 697, 1842

\bibitem[{{Kodama} {et~al.}(2004){Kodama}, {Balogh}, {Smail}, {Bower}, \&
  {Nakata}}]{Kodama+04}
{Kodama}, T., {Balogh}, M.~L., {Smail}, I., {Bower}, R.~G., \& {Nakata}, F.
  2004, \mnras, 354, 1103

\bibitem[{{Koyama} {et~al.}(2010){Koyama}, {Kodama}, {Shimasaku}, {Hayashi},
  {Okamura}, {Tanaka}, \& {Tokoku}}]{Koyama+10}
{Koyama}, Y., {Kodama}, T., {Shimasaku}, K., {et~al.} 2010, \mnras, 403, 1611

\bibitem[{{Koyama} {et~al.}(2008){Koyama}, {Kodama}, {Shimasaku}, {Okamura},
  {Tanaka}, {Lee}, {Im}, {Matsuhara}, {Takagi}, {Wada}, \& {Oyabu}}]{Koyama+08}
{Koyama}, Y., {Kodama}, T., {Shimasaku}, K., {et~al.} 2008, \mnras, 391, 1758

\bibitem[{{Krick} {et~al.}(2009){Krick}, {Surace}, {Thompson}, {Ashby}, {Hora},
  {Gorjian}, \& {Yan}}]{Krick+09}
{Krick}, J.~E., {Surace}, J.~A., {Thompson}, D., {et~al.} 2009, \apj, 700, 123

\bibitem[{{Kroupa}(2001)}]{Kroupa01}
{Kroupa}, P. 2001, \mnras, 322, 231

\bibitem[{{Lee} {et~al.}(2010){Lee}, {Le Floc'h}, {Sanders}, {Frayer},
  {Arnouts}, {Ilbert}, {Aussel}, {Salvato}, {Scoville}, \&
  {Kartaltepe}}]{Lee+10}
{Lee}, N., {Le Floc'h}, E., {Sanders}, D.~B., {et~al.} 2010, \apj, 717, 175

\bibitem[{{Makino} \& {Hut}(1997)}]{MH97}
{Makino}, J. \& {Hut}, P. 1997, \apj, 481, 83

\bibitem[{{Mamon}(1996)}]{Mamon96}
{Mamon}, G. 1996, in Third Paris Cosmology Colloquium, ed. H.~J. {de Vega} \&
  N.~{S{\'a}nchez}, 95, arXiv:astro-ph/9511101

\bibitem[{{Mamon} {et~al.}(2010){Mamon}, {Biviano}, \& {Murante}}]{Mamon+10}
{Mamon}, G.~A., {Biviano}, A., \& {Murante}, G. 2010, \aap, 520, A30

\bibitem[{{Maraston}(2005)}]{Maraston05}
{Maraston}, C. 2005, \mnras, 362, 799

\bibitem[{{Mauduit} \& {Mamon}(2007)}]{MM07}
{Mauduit}, J.-C. \& {Mamon}, G.~A. 2007, \aap, 475, 169

\bibitem[{{McQuinn} {et~al.}(2010){McQuinn}, {Skillman}, {Cannon}, {Dalcanton},
  {Dolphin}, {Hidalgo-Rodr{\'{\i}}guez}, {Holtzman}, {Stark}, {Weisz}, \&
  {Williams}}]{McQuinn+10}
{McQuinn}, K.~B.~W., {Skillman}, E.~D., {Cannon}, J.~M., {et~al.} 2010, \apj,
  721, 297

\bibitem[{{Merluzzi} {et~al.}(2010){Merluzzi}, {Mercurio}, {Haines}, {Smith},
  {Busarello}, \& {Lucey}}]{Merluzzi+10}
{Merluzzi}, P., {Mercurio}, A., {Haines}, C.~P., {et~al.} 2010, \mnras, 402,
  753

\bibitem[{{Metcalfe} {et~al.}(2005){Metcalfe}, {Fadda}, \& {Biviano}}]{MFB05}
{Metcalfe}, L., {Fadda}, D., \& {Biviano}, A. 2005, Space Science Reviews, 119,
  425

\bibitem[{{Metcalfe} {et~al.}(2003){Metcalfe}, {Kneib}, {McBreen}, {Altieri},
  {Biviano}, {Delaney}, {Elbaz}, {Kessler}, {Leech}, {Okumura}, {Ott},
  {Perez-Martinez}, {Sanchez-Fernandez}, \& {Schulz}}]{Metcalfe+03}
{Metcalfe}, L., {Kneib}, J.-P., {McBreen}, B., {et~al.} 2003, \aap, 407, 791

\bibitem[{{Meusinger} {et~al.}(2000){Meusinger}, {Brunzendorf}, \&
  {Krieg}}]{MBK00}
{Meusinger}, H., {Brunzendorf}, J., \& {Krieg}, R. 2000, \aap, 363, 933

\bibitem[{{Moore} {et~al.}(1996){Moore}, {Katz}, {Lake}, {Dressler}, \&
  {Oemler}}]{Moore+96}
{Moore}, B., {Katz}, N., {Lake}, G., {Dressler}, A., \& {Oemler}, Jr., A. 1996,
  \nat, 379, 613

\bibitem[{{Murakami} {et~al.}(2007){Murakami}, {Baba}, {Barthel}, {Clements},
  {Cohen}, {Doi}, {Enya}, {Figueredo}, {Fujishiro}, {Fujiwara}, {Fujiwara},
  {Garcia-Lario}, {Goto}, {Hasegawa}, {Hibi}, {Hirao}, {Hiromoto}, {Hong},
  {Imai}, {Ishigaki}, {Ishiguro}, {Ishihara}, {Ita}, {Jeong}, {Jeong},
  {Kaneda}, {Kataza}, {Kawada}, {Kawai}, {Kawamura}, {Kessler}, {Kester},
  {Kii}, {Kim}, {Kim}, {Kobayashi}, {Koo}, {Kwon}, {Lee}, {Lorente}, {Makiuti},
  {Matsuhara}, {Matsumoto}, {Matsuo}, {Matsuura}, {M{\"u}ller}, {Murakami},
  {Nagata}, {Nakagawa}, {Naoi}, {Narita}, {Noda}, {Oh}, {Ohnishi}, {Ohyama},
  {Okada}, {Okuda}, {Oliver}, {Onaka}, {Ootsubo}, {Oyabu}, {Pak}, {Park},
  {Pearson}, {Rowan-Robinson}, {Saito}, {Sakon}, {Salama}, {Sato}, {Savage},
  {Serjeant}, {Shibai}, {Shirahata}, {Sohn}, {Suzuki}, {Takagi}, {Takahashi},
  {Tanab{\'e}}, {Takeuchi}, {Takita}, {Thomson}, {Uemizu}, {Ueno}, {Usui},
  {Verdugo}, {Wada}, {Wang}, {Watabe}, {Watarai}, {White}, {Yamamura},
  {Yamauchi}, \& {Yasuda}}]{Murakami+07}
{Murakami}, H., {Baba}, H., {Barthel}, P., {et~al.} 2007, \pasj, 59, 369

\bibitem[{{Navarro} {et~al.}(1997){Navarro}, {Frenk}, \& {White}}]{NFW97}
{Navarro}, J.~F., {Frenk}, C.~S., \& {White}, S. D.~M. 1997, \apj, 490, 493

\bibitem[{{Negroponte} \& {White}(1983)}]{NW83}
{Negroponte}, J. \& {White}, S.~D.~M. 1983, \mnras, 205, 1009

\bibitem[{{Oemler} {et~al.}(2009){Oemler}, {Dressler}, {Kelson}, {Rigby},
  {Poggianti}, {Fritz}, {Morrison}, \& {Smail}}]{Oemler+09}
{Oemler}, A., {Dressler}, A., {Kelson}, D., {et~al.} 2009, \apj, 693, 152

\bibitem[{{Oliver} {et~al.}(2010){Oliver}, {Frost}, {Farrah},
  {Gonzalez-Solares}, {Shupe}, {Henriques}, {Roseboom}, {Alfonso-Luis},
  {Babbedge}, {Frayer}, {Lencz}, {Lonsdale}, {Masci}, {Padgett}, {Polletta},
  {Rowan-Robinson}, {Siana}, {Smith}, {Surace}, \& {Vaccari}}]{Oliver+10}
{Oliver}, S., {Frost}, M., {Farrah}, D., {et~al.} 2010, \mnras, 405, 2279

\bibitem[{{Oyaizu} {et~al.}(2008){Oyaizu}, {Lima}, {Cunha}, {Lin}, {Frieman},
  \& {Sheldon}}]{Oyaizu+08}
{Oyaizu}, H., {Lima}, M., {Cunha}, C.~E., {et~al.} 2008, \apj, 674, 768

\bibitem[{{Peacock}(1983)}]{Peacock83}
{Peacock}, J.~A. 1983, \mnras, 202, 615

\bibitem[{{Peng} {et~al.}(2010){Peng}, {Lilly}, {Kova{\v c}}, {Bolzonella},
  {Pozzetti}, {Renzini}, {Zamorani}, {Ilbert}, {Knobel}, {Iovino}, {Maier},
  {Cucciati}, {Tasca}, {Carollo}, {Silverman}, {Kampczyk}, {de Ravel},
  {Sanders}, {Scoville}, {Contini}, {Mainieri}, {Scodeggio}, {Kneib}, {Le
  F{\`e}vre}, {Bardelli}, {Bongiorno}, {Caputi}, {Coppa}, {de la Torre},
  {Franzetti}, {Garilli}, {Lamareille}, {Le Borgne}, {Le Brun}, {Mignoli},
  {Perez Montero}, {Pello}, {Ricciardelli}, {Tanaka}, {Tresse}, {Vergani},
  {Welikala}, {Zucca}, {Oesch}, {Abbas}, {Barnes}, {Bordoloi}, {Bottini},
  {Cappi}, {Cassata}, {Cimatti}, {Fumana}, {Hasinger}, {Koekemoer},
  {Leauthaud}, {Maccagni}, {Marinoni}, {McCracken}, {Memeo}, {Meneux}, {Nair},
  {Porciani}, {Presotto}, \& {Scaramella}}]{Peng+10}
{Peng}, Y., {Lilly}, S.~J., {Kova{\v c}}, K., {et~al.} 2010, \apj, 721, 193

\bibitem[{{Pereira} {et~al.}(2010){Pereira}, {Haines}, {Smith}, {Egami},
  {Moran}, {Finoguenov}, {Hardegree-Ullman}, {Okabe}, {Rawle}, \&
  {Rex}}]{Pereira+10}
{Pereira}, M.~J., {Haines}, C.~P., {Smith}, G.~P., {et~al.} 2010, \aap, 518,
  L40

\bibitem[{{Pilbratt} {et~al.}(2010){Pilbratt}, {Riedinger}, {Passvogel},
  {Crone}, {Doyle}, {Gageur}, {Heras}, {Jewell}, {Metcalfe}, {Ott}, \&
  {Schmidt}}]{Pilbratt+10}
{Pilbratt}, G.~L., {Riedinger}, J.~R., {Passvogel}, T., {et~al.} 2010, \aap,
  518, L1

\bibitem[{{Polletta} {et~al.}(2007){Polletta}, {Tajer}, {Maraschi},
  {Trinchieri}, {Lonsdale}, {Chiappetti}, {Andreon}, {Pierre}, {Le F{\`e}vre},
  {Zamorani}, {Maccagni}, {Garcet}, {Surdej}, {Franceschini}, {Alloin},
  {Shupe}, {Surace}, {Fang}, {Rowan-Robinson}, {Smith}, \&
  {Tresse}}]{Polletta+07}
{Polletta}, M., {Tajer}, M., {Maraschi}, L., {et~al.} 2007, \apj, 663, 81

\bibitem[{{Porter} \& {Raychaudhury}(2007)}]{PR07}
{Porter}, S.~C. \& {Raychaudhury}, S. 2007, \mnras, 375, 1409

\bibitem[{{Porter} {et~al.}(2008){Porter}, {Raychaudhury}, {Pimbblet}, \&
  {Drinkwater}}]{Porter+08}
{Porter}, S.~C., {Raychaudhury}, S., {Pimbblet}, K.~A., \& {Drinkwater}, M.~J.
  2008, \mnras, 388, 1152

\bibitem[{{Quilis} {et~al.}(2000){Quilis}, {Moore}, \& {Bower}}]{QMB00}
{Quilis}, V., {Moore}, B., \& {Bower}, R. 2000, Science, 288, 1617

\bibitem[{{Quintana} {et~al.}(1996){Quintana}, {Ramirez}, \& {Way}}]{QRW96}
{Quintana}, H., {Ramirez}, A., \& {Way}, M.~J. 1996, \aj, 112, 36

\bibitem[{{Rawle} {et~al.}(2010){Rawle}, {Chung}, {Fadda}, {Rex}, {Egami},
  {P{\'e}rez-Gonz{\'a}lez}, {Altieri}, {Blain}, {Bridge}, {Fiedler},
  {Gonzalez}, {Pereira}, {Richard}, {Smail}, {Valtchanov}, {Zemcov},
  {Appleton}, {Bock}, {Boone}, {Clement}, {Combes}, {Dowell},
  {Dessauges-Zavadsky}, {Ilbert}, {Ivison}, {Jauzac}, {Kneib}, {Lutz},
  {Pell{\'o}}, {Rieke}, {Rodighiero}, {Schaerer}, {Smith}, {Walth}, {van der
  Werf}, \& {Werner}}]{Rawle+10}
{Rawle}, T.~D., {Chung}, S.~M., {Fadda}, D., {et~al.} 2010, arXiv:1005.3822

\bibitem[{{Rex} {et~al.}(2010){Rex}, {Rawle}, {Egami},
  {P{\'e}rez-Gonz{\'a}lez}, {Zemcov}, {Aretxaga}, {Chung}, {Fadda}, {Gonzalez},
  {Hughes}, {Horellou}, {Johansson}, {Kneib}, {Richard}, {Altieri}, {Fiedler},
  {Pereira}, {Rieke}, {Smail}, {Valtchanov}, {Blain}, {Bock}, {Boone},
  {Bridge}, {Clement}, {Combes}, {Dowell}, {Dessauges-Zavadsky}, {Ilbert},
  {Ivison}, {Jauzac}, {Lutz}, {Omont}, {Pell{\'o}}, {Rodighiero}, {Schaerer},
  {Smith}, {Walth}, {van der Werf}, {Werner}, {Austermann}, {Ezawa}, {Kawabe},
  {Kohno}, {Perera}, {Scott}, {Wilson}, \& {Yun}}]{Rex+10}
{Rex}, M., {Rawle}, T.~D., {Egami}, E., {et~al.} 2010, \aap, 518, L13+

\bibitem[{{Rieke} {et~al.}(2009){Rieke}, {Alonso-Herrero}, {Weiner},
  {P{\'e}rez-Gonz{\'a}lez}, {Blaylock}, {Donley}, \& {Marcillac}}]{Rieke+09}
{Rieke}, G.~H., {Alonso-Herrero}, A., {Weiner}, B.~J., {et~al.} 2009, \apj,
  692, 556

\bibitem[{{Rujopakarn} {et~al.}(2010){Rujopakarn}, {Eisenstein}, {Rieke},
  {Papovich}, {Cool}, {Moustakas}, {Jannuzi}, {Kochanek}, {Rieke}, {Dey},
  {Eisenhardt}, {Murray}, {Brown}, \& {Le Floc'h}}]{Rujopakarn+10}
{Rujopakarn}, W., {Eisenstein}, D.~J., {Rieke}, G.~H., {et~al.} 2010, \apj,
  718, 1171

\bibitem[{{Saintonge} {et~al.}(2008){Saintonge}, {Tran}, \& {Holden}}]{STH08}
{Saintonge}, A., {Tran}, K.-V.~H., \& {Holden}, B.~P. 2008, \apjl, 685, L113

\bibitem[{{Sanders} {et~al.}(1988){Sanders}, {Soifer}, {Elias}, {Madore},
  {Matthews}, {Neugebauer}, \& {Scoville}}]{Sanders+88}
{Sanders}, D.~B., {Soifer}, B.~T., {Elias}, J.~H., {et~al.} 1988, \apj, 325, 74

\bibitem[{{Schechter}(1976)}]{Schechter76}
{Schechter}, P. 1976, \apj, 203, 297

\bibitem[{{Silva} {et~al.}(1998){Silva}, {Granato}, {Bressan}, \&
  {Danese}}]{Silva+98}
{Silva}, L., {Granato}, G.~L., {Bressan}, A., \& {Danese}, L. 1998, \apj, 509,
  103

\bibitem[{{Spitzer} \& {Baade}(1951)}]{SB51}
{Spitzer}, L.~J. \& {Baade}, W. 1951, \apj, 113, 413

\bibitem[{{Temporin} {et~al.}(2009){Temporin}, {Duc}, {Ilbert}, \&
  {XMM-LSS/SWIRE collaboration}}]{TDI09}
{Temporin}, S., {Duc}, P., {Ilbert}, O., \& {XMM-LSS/SWIRE collaboration}.
  2009, Astronomische Nachrichten, 330, 915

\bibitem[{{Teyssier} {et~al.}(2010){Teyssier}, {Chapon}, \& {Bournaud}}]{TCB10}
{Teyssier}, R., {Chapon}, D., \& {Bournaud}, F. 2010, \apjl, 720, L149

\bibitem[{{Tran} {et~al.}(2009){Tran}, {Saintonge}, {Moustakas}, {Bai},
  {Gonzalez}, {Holden}, {Zaritsky}, \& {Kautsch}}]{Tran+09}
{Tran}, K., {Saintonge}, A., {Moustakas}, J., {et~al.} 2009, \apj, 705, 809

\bibitem[{{Umemura} {et~al.}(1999){Umemura}, {Fukue}, \& {Mineshige}}]{UFM99}
{Umemura}, M., {Fukue}, J., \& {Mineshige}, S. 1999, Advances in Space
  Research, 23, 1095

\bibitem[{{van den Bosch}(2002)}]{vandenBosch02}
{van den Bosch}, F.~C. 2002, \mnras, 331, 98

\bibitem[{{Werner} {et~al.}(2004){Werner}, {Roellig}, {Low}, {Rieke}, {Rieke},
  {Hoffmann}, {Young}, {Houck}, {Brandl}, {Fazio}, {Hora}, {Gehrz}, {Helou},
  {Soifer}, {Stauffer}, {Keene}, {Eisenhardt}, {Gallagher}, {Gautier}, {Irace},
  {Lawrence}, {Simmons}, {Van Cleve}, {Jura}, {Wright}, \&
  {Cruikshank}}]{Werner+04}
{Werner}, M.~W., {Roellig}, T.~L., {Low}, F.~J., {et~al.} 2004, \apjs, 154, 1

\bibitem[{{Wild} {et~al.}(2009){Wild}, {Walcher}, {Johansson}, {Tresse},
  {Charlot}, {Pollo}, {Le F{\`e}vre}, \& {de Ravel}}]{Wild+09}
{Wild}, V., {Walcher}, C.~J., {Johansson}, P.~H., {et~al.} 2009, \mnras, 395,
  144

\bibitem[{{Wojtak} {et~al.}(2007){Wojtak}, {{\L}okas}, {Mamon},
  {Gottl{\"o}ber}, {Prada}, \& {Moles}}]{Wojtak+07}
{Wojtak}, R., {{\L}okas}, E.~L., {Mamon}, G.~A., {et~al.} 2007, \aap, 466, 437

\bibitem[{{Yahil} \& {Vidal}(1977)}]{YV77}
{Yahil}, A. \& {Vidal}, N.~V. 1977, \apj, 214, 347

\bibitem[{{Younger} {et~al.}(2009){Younger}, {Hayward}, {Narayanan}, {Cox},
  {Hernquist}, \& {Jonsson}}]{Younger+09}
{Younger}, J.~D., {Hayward}, C.~C., {Narayanan}, D., {et~al.} 2009, \mnras,
  396, L66

\bibitem[{{Zucca} {et~al.}(2009){Zucca}, {Bardelli}, {Bolzonella}, {Zamorani},
  {Ilbert}, {Pozzetti}, {Mignoli}, {Kova{\v c}}, {Lilly}, {Tresse}, {Tasca},
  {Cassata}, {Halliday}, {Vergani}, {Caputi}, {Carollo}, {Contini}, {Kneib},
  {Le F{\`e}vre}, {Mainieri}, {Renzini}, {Scodeggio}, {Bongiorno}, {Coppa},
  {Cucciati}, {de La Torre}, {de Ravel}, {Franzetti}, {Garilli}, {Iovino},
  {Kampczyk}, {Knobel}, {Lamareille}, {Le Borgne}, {Le Brun}, {Maier},
  {Pell{\`o}}, {Peng}, {Perez-Montero}, {Ricciardelli}, {Silverman}, {Tanaka},
  {Abbas}, {Bottini}, {Cappi}, {Cimatti}, {Guzzo}, {Koekemoer}, {Leauthaud},
  {Maccagni}, {Marinoni}, {McCracken}, {Memeo}, {Meneux}, {Moresco}, {Oesch},
  {Porciani}, {Scaramella}, {Arnouts}, {Aussel}, {Capak}, {Kartaltepe},
  {Salvato}, {Sanders}, {Scoville}, {Taniguchi}, \& {Thompson}}]{Zucca+09}
{Zucca}, E., {Bardelli}, S., {Bolzonella}, M., {et~al.} 2009, \aap, 508, 1217

\end{thebibliography}

\end{document}